\documentclass[reprint,amsmath,amssymb,aps,superscriptaddress,twocolumn,nofootinbib,10pt]{revtex4-1}
\usepackage[colorlinks=true,allcolors=blue]{hyperref}
\usepackage{graphicx,color}
\usepackage{dcolumn}
\usepackage{subfigure}
\usepackage{bm}
\usepackage{amsmath,amsfonts,amssymb}
\usepackage{acronym}
\usepackage{enumitem}
\usepackage{array}
\usepackage{multirow}
\usepackage{CJK}
\allowdisplaybreaks
\newcommand{\be}{\begin{equation}}
\newcommand{\ee}{\end{equation}}
\newcommand{\bea}{\begin{eqnarray}}
\newcommand{\eea}{\end{eqnarray}}

\newcommand{\SOUTHCUT}{School of Physics and Optoelectronics, South China University of Technology, Guangzhou 510641, People's Republic of China}
\newcommand{\NCUa}{Department of physics, Nanchang University, Nanchang, 330031, China}
\newcommand{\NCUb}{Center for Relativistic Astrophysics and High Energy Physics, Nanchang University, Nanchang, 330031, China}

\newacro{EMRI}{extreme mass-ratio inspirals}
\newacro{MBH}{massive black hole}
\newacro{BH}{black hole}
\newacro{GR}{general relativity}
\newacro{HKBH}{hairy Kerr black hole}
\newacro{KNBH}{Kerr-Newmann black hole}
\newacro{KBH}{Kerr black hole}
\newacro{NHT}{no-hair theorem}
\newacro{DWD}{double white dwarf}
\newacro{GW}{gravitational wave}
\newacro{AK}{analytic kludge}
\newacro{NK}{numerical kludge}
\newacro{AAK}{augmented analytic kludge}
\newacro{CO}{compact object}
\newacro{PE}{parameter estimation}
\newacro{SNR}{signal-to-noise ratio}
\newacro{PN}{post newtonion}
\newacro{FIM}{Fisher information matrix}
\newacro{LSO}{last stable orbit}
\newacro{ISCO}{innermost stable circular orbit}
\newacro{BBH}{Binary Black Hole}
\newacro{BNS}{Binary Neutron Star}
\newacro{NS}{Neutron Star}
\newacro{KN}{Kerr-Newmann}
\newcommand{\beq}{\begin{equation}}
\newcommand{\eeq}{\end{equation}}
\newcommand{\beqa}{\begin{eqnarray}}
\newcommand{\eeqa}{\end{eqnarray}}

\newcommand{\Op}{\hat{\mathcal{D}}_2}

\newcommand{\afn}{u}

\def\lsim{\mathrel{\rlap{\lower4pt\hbox{\hskip0.5pt$\sim$}}
    \raise1pt\hbox{$<$}}}         %less than or approx. symbol
\def\gsim{\mathrel{\rlap{\lower4pt\hbox{\hskip0.5pt$\sim$}}
    \raise1pt\hbox{$>$}}}         %greater than or approx. symbol
\begin{document}
\begin{CJK*}{UTF8}{gbsn}
\title{{\Large {\bf Eccentric extreme-mass-ratio inspirals: a new window into ultra-light vector fields}}}

\author{Tieguang Zi}
\email{zitieguang@ncu.edu.cn}
\affiliation{\NCUa}
\affiliation{\NCUb}
\affiliation{\SOUTHCUT}
\author{Fu-Wen Shu}
\email{shufuwen@ncu.edu.cn}
\affiliation{\NCUa}
\affiliation{\NCUb}

\begin{abstract}
Space-based gravitational-wave detectors, such as the Laser Interferometer Space Antenna (LISA), provide a platform to probe new fundamental fields through extreme-mass-ratio inspirals (EMRIs), where a compact secondary object carrying scalar or vector charges inspirals into a massive primary. In a theory-agnostic framework, we compute the ultra-light vector and gravitational radiation emitted by eccentric EMRIs and determine the corresponding inspiral trajectories. We evaluate the impact of a massive vector (Proca) field on EMRIs waveform through dephasing and mismatches with predictions by general relativity. Using a Fisher information matrix analysis, we further assess LISA's capability to constrain the Proca mass from future EMRIs observations. We find that orbital eccentricity can improve  estimation accuracy of parameters, making the vector mass $\mu$ become detectable for the case of $\mu=0.02$ . Correlation analysis further reveals strong positive dependencies between the Proca mass and intrinsic source parameters, indicating that improved measurement of these parameters directly tightens constraints on vector mass. These results demonstrate that high-eccentricity EMRIs observed by LISA offer a powerful channel to detect or constrain massive vector-field extensions of GR in the strong-field regime.
\end{abstract}
\maketitle
\end{CJK*}

\begin{abstract}
Space-based gravitational-wave detectors such as the Laser Interferometer Space Antenna (LISA) provide a promising platform to probe new fundamental fields through extreme mass-ratio inspirals (EMRIs), in which a compact secondary object carrying scalar or vector charges inspirals into a massive primary black hole. Within a theory-agnostic framework, we compute the ultra-light vector and gravitational radiation emitted by eccentric EMRIs and determine the associated inspiral trajectories. We investigate the impact of a massive vector (Proca) field on EMRI waveforms through phase dephasing and mismatches with respect to general relativity. Using a Fisher information matrix analysis, we evaluate LISA¡¯s capability to constrain the Proca mass from future EMRI observations. Our results show that orbital eccentricity significantly enhances parameter estimation, reducing the relative uncertainty in the Proca mass from $774\%$ to $121\%$ for $\mu = 0.01$, and achieving a precision of $6.54\%$ ($\sim 8.7\times10^{-18}\,\mathrm{eV}$) for $\mu = 0.02$. Correlation analysis reveals strong positive dependencies between the Proca mass and several intrinsic parameters, indicating that improved measurement of these parameters directly tightens constraints on $\mu$. Numerical stability tests of the covariance matrix confirm that more than $90\%$ of realizations yield relative errors below $0.3\%$, ensuring robustness of the parameter estimation. These findings demonstrate that high-eccentricity EMRIs observed by LISA can serve as an effective probe of ultra-light vector fields and potential deviations from general relativity in the strong-field regime.
\end{abstract}

\section{Introduction}
Over the past decade, observations of gravitational-wave (GW) events with ground-based interferometers have greatly deepened understanding of the Universe and the nature of compact objects,
enabling to test the gravitational theories in the strong field region
\cite{LIGOScientific:2016aoc,KAGRA:2021vkt,LIGOScientific:2020ibl,LIGOScientific:2016lio,LIGOScientific:2021sio,LIGOScientific:2018dkp,LIGOScientific:2020tif,Wang:2023jah}. However, these GW events have the smaller signal-to-noise ratios (SNR), searches for deviations from standard general relativity (GR) in the LIGO-Virgo-KAGRA (LVK) data stream have thus far found no convincing evidence of such deviations.
The future GW detectors including Einstein Telescope (ET) \cite{Punturo:2010zz}, LISA \cite{LISA:2017pwj}, TianQin \cite{TianQin:2015yph, TianQin:2020hid} and Taiji \cite{Ruan:2018tsw}
would extend sensitivity across a broader mass range and achieve much higher SNRs,
offering a valuable opportunity to constrain fundamental theories with an extraordinary precision~\cite{Bian:2021ini,Luo:2025ewp,Bian:2025ifp} and to search for new degrees of freedom, such as ultra-light scalar and vector fields~\cite{Chagoya:2016aar,Minamitsuji:2016ydr,Babichev:2017rti,Herdeiro:2016tmi,Heisenberg:2017xda,Kase:2018owh,Kase:2018voo}.

Modifications to GR are motivated by cosmological puzzles such as dark matter and dark energy, as well as the quest for a consistent quantum theory of gravity and extensions of the Standard Model~\cite{Berti:2015itd,Barack:2018yly,Barausse:2020rsu,Sotiriou:2014yhm}. Many of these theories naturally introduce additional fields mediating the gravitational interaction. The effects of such fields may be revealed in the dynamics of compact objects in the strong-curvature regime, where dipole radiation and energy loss beyond the GR prediction can significantly affect the phase evolution of GWs. Observationally, stringent constraints have been placed on massless scalar fields from pulsar timing~\cite{Antoniadis:2013pzd,Anderson:2019eay}, while GW searches for scalar-induced dipole radiation in binary mergers have found no evidence of deviations from GR~\cite{Yagi:2012gp,Nair:2019iur,Perkins:2021mhb,Sanger:2024axs,Julie:2024fwy,Lyu:2022gdr,Wang:2023wgv,Saffer:2021gak}. Nevertheless, it has been argued that LISA-like detectors could detect such new fields with higher precision using the asymmetric binaries with long inspiral timescales~\cite{Maselli:2020zgv,Maselli:2021men,Barsanti:2022ana,Barsanti:2022vvl,DellaRocca:2024pnm,Barsanti:2024kul,Speri:2024qak}.

A particularly promising class of sources for LISA are extreme-mass-ratio inspirals (EMRIs), in which a stellar-mass compact object (secondary) spirals into a massive black hole (MBH, primary) in a galactic nucleus. Due to their extreme mass ratios $\eta \leq 10^{-4}$, EMRI signals can last several years in the LISA band~\cite{Berry:2019wgg,LISA:2022yao}, accumulating $10^4-10^5$ orbital cycles in the strong-field region. Their orbital dynamics, often involving significant eccentricity and inclination, are expected to encode detailed information about the MBH spacetime~\cite{Barack:2018yly,Barack:2006pq,Babak:2017tow,Fan:2020zhy,Zi:2021pdp,LISA:2022kgy}. This makes EMRIs excellent laboratories to probe both deviations from GR~\cite{Dolan:2007mj,Arvanitaki:2016qwi,Arvanitaki:2014wva,Pani:2012vp,Pani:2012bp,Witek:2012tr,East:2017ovw,East:2018glu,Dolan:2018dqv,Siemonsen:2019ebd} and environmental effects surrounding MBHs~\cite{Barausse:2014tra,Hannuksela:2018izj,Cardoso:2019upw,Toubiana:2020drf,DeLuca:2022xlz,Cole:2022yzw,Dyson:2025dlj,Becker:2024ibd,Pan:2021ksp,Zwick:2025wkt,Kejriwal:2023djc}. However, accurately modeling eccentric EMRI waveforms in the presence of such effects remains highly challenging.
On the other hand, importantly, EMRIs also provide a natural arena for testing new fundamental fields, including scalar~\cite{Maselli:2020zgv,Maselli:2021men,Barsanti:2022ana,Barsanti:2022vvl,DellaRocca:2024pnm,Barsanti:2024kul,Speri:2024qak,Jiang:2021htl,Guo:2022euk,Zhang:2022rfr}, vector~\cite{Zhang:2022hbt,Zhang:2024ogc,Zi:2024lmt}, and tensor fields~\cite{Cardoso:2018zhm}. Additional energy flux induced by such fields can imprint distinct signatures on the GW phase, thereby allowing stringent constraints on their mass and coupling. Consequently, the study of EMRIs not only improves our understanding of the fundamental structure of spacetime near MBHs but also offers a unique probe of new physics.

The no-hair theorem implies that stationary black holes in many scalar-tensor theories reduce to the Kerr metric without scalar charge~\cite{Chase:1970omy,Bekenstein:1995un,Hui:2012qt,Sotiriou:2011dz}.
In general, the stationary BHs in scalar-tensor theories carry the non-trivial scalar hair, the extra scalar charge is related with the interaction between scalar filed and higher-order curvature invariant terms \cite{Kanti:1995vq,Alexander:2009tp,Yunes:2011we}.
Specifically, the deviation from BHs in GR is approximately inverse the powers of their mass. In other words, the scalar charge carried with MBHs would become negligible due to the smaller curvature near the horizon. Whereas, secondary objects in EMRIs may evade no-hair theorem and carry nontrivial scalar or vector charges due to their smaller mass and higher curvature~\cite{Kanti:1995vq,Alexander:2009tp,Yunes:2011we}. Motivated by this, we focus on EMRIs in the Einstein-Proca theory, where a massive vector field couples non-minimally to gravity~\cite{Chagoya:2016aar,Minamitsuji:2016ydr,Babichev:2017rti,Herdeiro:2016tmi,Heisenberg:2017xda,
Kase:2018owh,Kase:2018voo,Minamitsuji:2017pdr,Kase:2017egk,Kase:2020yhw}. In this framework, a secondary object endowed with Proca hair inspirals along eccentric orbits around a Schwarzschild MBH, extending previous studies restricted to circular orbits~\cite{Zi:2024lmt}. The resulting GW signals encode information about the Proca field, providing a novel opportunity to probe deviations from GR within the LISA band~\cite{Maselli:2020zgv,Maselli:2021men,Konoplya:2005hr,Konoplya:2006gq,Zi:2024lmt,Lutfuoglu:2025qkt}.
The detectability of such effects of fundamental fields depends on the Compton wavelength of the Proca field being comparable to the binary separation or compact-object size. In geometric units $(G=c=1)$, the relevant condition is $\mu_v M \lesssim 1$, here $M$ is the MBH mass, and $\mu_v$ the Proca mass, which can be expressed in electronvolts,
\begin{align}
\mu_v[{\rm eV}] &\sim \left(\frac{\mu_v M}{0.75}\right) \left(\frac{10^6 M_\odot}{M}\right) 10^{-16} \rm eV
\nonumber \\   & = \left(\frac{\mu}{0.75}\right) \left(\frac{10^6 M_\odot}{M}\right) 10^{-16} \rm eV\;,
\end{align}
where $\mu \equiv \mu_v M$ is a dimensionless quantity and $M_\odot$ is the solar mass~\cite{Brito:2015oca}.

In this paper, we investigate the modification of EMRI fluxes in the Einstein-Proca theory using perturbation methods. We compute the vector and gravitational fluxes from eccentric equatorial orbits, evolve the inspiral trajectories adiabatically, and compute waveforms via the quadrupole formula. By comparing Proca-induced radiation with GR and massless-vector cases, we assess the resulting waveform dephasing and mismatches. The paper is organized as follows. In Sec.\ref{env:mehod}, we present the theoretical framework and computational procedure. In Sec.\ref{result}, we report the results, including waveform comparisons, dephasing, and mismatches. Finally, Sec.~\ref{conslusion} summarizes our findings.

%%%%%%%%%%%%%%%%%%%%%%%%%%%%%%%%%%%%%%%%%%%%%%%%%%%%%%%%%%%%%%%%%%%%%%%%%%%%%%%%%%%%%%%%%%%%%%%%%%%%%%
\section{Theoretical model} \label{env:mehod}
\subsection{Action}
In this section, we briefly review  a theoretical framework describing EMRIs system with secondary haired Proca charge~\cite{Ramazanoglu:2016kul,Maselli:2020zgv,Maselli:2021men,Barsanti:2022vvl,Zi:2024lmt}.
We consider a theory with the Proca field $A$ nonminimally coupled to a metric $g$, the general action that describes the theory can be written as follows \cite{Konoplya:2005hr,Konoplya:2006gq,Chagoya:2016aar,Minamitsuji:2016ydr,Babichev:2017rti,Herdeiro:2016tmi,Heisenberg:2017xda,Kase:2018owh,Kase:2018voo}
\begin{equation}\label{action:general}
S[g,A,\Psi] = S_0[g,A]+ \alpha S_c[g, \Psi] + S_m[g,A,\Psi],
\end{equation}
where $S_0[g,A]$ is related with the Einstein-Hilbert action and the kinetic term for the Proca field
\begin{equation}\label{action:S0}
S_0[g,A] = \int d^4x \frac{\sqrt{-g}}{16\pi} \left(R-\mathcal{L} \right),
\end{equation}
with
\begin{equation}
\mathcal{L} = \frac{1}{4}F_{\mu\nu}F^{\mu\nu}+ \frac{\mu_v^2}{2}A_{\mu}A^\mu,
\end{equation}
$R$ is the Ricci scalar determined with metric $g$, $F_{\mu\nu}=\nabla_\mu A_{\nu}- \nabla_\nu A_{\mu}$ is the Proca field strength and $\mu_v$ is the
mass of Proca field. The quantify $S_m[g,A,\Psi]$ denotes to the matter fields section. $\alpha S_c$ describes the non-minimal coupling between the metric $g$ and Proca field $A$, where the coupling constant $\alpha$ has a dimension $[\alpha]=(\rm mass)^n$ with a  positive integer $n$.
Because the scale of secondary is far less than the typical scale of spacetime of MBH, we can take the orbiting object as a point particle using the ideology of skeletonized approach introduced in Refs.~\cite{Ramazanoglu:2016kul,Maselli:2020zgv,Maselli:2021men,Barsanti:2022vvl}.
Specifically, the matter field action $S_m$ can be replaced with the action $S_p$ of particle, then we ignore the coupling of secondary and Proca field, and the particle's action $S_p$  with a mass $m_p$
\begin{eqnarray}\label{action:Sp}
S_p(g,A,\Psi) &=& -\int m_p d\tau + \int d^4x \sqrt{-g} A_\mu J^\mu ,
%&=&-m_p \int d\tau \sqrt{-g_{\mu\nu} \frac{dx^\mu}{d\tau} \frac{dx^\nu}{d\tau} }
\end{eqnarray}
where the current density for vector field is
$J^\mu = q m_p \int d\tau \frac{dy^\mu}{d\tau} \frac{\delta^{(4)} (x-y(\tau))}{\sqrt{-g}}$, $y(\tau)$ is the worldline of secondary with a proper time $\tau$, and $q$ is the vector charge carried by secondary object.

As the BH solutions from action \eqref{action:general} can continuously return to the BHs predicted by GR when the constant continuously approaches zero $(\alpha \rightarrow0)$ \cite{Garcia-Saenz:2021uyv}, the deviations from the cases of GR rely on the dimensionless parameter
\begin{equation}
\xi = \frac{\alpha}{M^n} = \eta^n\frac{\alpha}{m_p^n}.
\end{equation}
According to the constraint of astrophysical observation, the parameter $\alpha/m_p^n$
should be order of magnitudes of $\mathcal{O}(1)$, even less than one \cite{Nair:2019iur}, implying that $\xi$ is order of $\eta^n$. Hence, we can claim that the MBHs from action \eqref{action:general} can be approximately described by the Schwarzchild metric because the parameter $\xi ~(\xi\ll1)$ controls the deviation of BHs solutions.
In the special case $\alpha=0$, the theory reduces to GR and the no-hair theorem holds exactly.

Therefore, for EMRIs in a class of theories involving light fields, one may assume that the spacetime of the primary is described by a Schwarzschild metric up to corrections of order $\mathcal{O}(\eta^n \xi)$. The radiation from the inspiraling secondary can then be treated perturbatively in this Schwarzschild background~\cite{Zi:2024lmt}. However, if black hole solutions in the modified theory deviate strongly from the no-hair theorem, the primary object in EMRIs cannot be approximated by a Schwarzschild or Kerr black hole. The present framework should thus be regarded as a first step, providing a rough estimate of the impact of the Proca field on EMRI fluxes.

\subsection{Field equations}
From the action in Eq.~\eqref{action:general},  we can obtain gravitational and Proca field equations using variation with respect to metric and vector
\begin{eqnarray}\label{eq:fields}
&G^{\mu\nu} &= \mathcal{T}^{\mu\nu}_{\rm Proca}  +  \mathcal{T}_p^{\mu\nu} + \alpha \mathcal{T}^{\mu\nu}_c \,, \nonumber \\
&\nabla_\rho F^{\rho\mu} &-\mu_v^2 A^\mu+ \frac{8\pi\alpha}{\sqrt{-g}} \frac{\delta S_c}{\delta A^\mu} = 8\pi J^\mu \;,
\end{eqnarray}
where $\mathcal{T}^{\rm Proca}_{\mu\nu}$ is the stress-energy tensor of the Proca field
\begin{widetext}
\begin{align}
\mathcal{T}_{\rm Proca}^{\mu\nu}  =  8\pi \big(-\frac{1}{4} F_{\rho\sigma} F^{\rho\sigma} g^{\mu\nu} + F_{~\rho}^{\mu} F^{\rho\nu}  - \frac{1}{2} \mu_v^2 g^{\mu\nu}A_\rho A^\rho
+\mu_v^2 A^\mu A^\nu \big),
\end{align}
\end{widetext}
and $\mathcal{T}^{c}_{\mu\nu}=- \frac{16\pi}{\sqrt{-g}} \frac{\delta S_c}{\delta g^{\mu\nu}}$.
The stress-energy tensor of smaller object can be obtained by varying action $S_p$ in Eq.~\eqref{action:Sp}
\begin{align}
\mathcal{T}_p^{\mu\nu} &= 8\pi m_p \int d\tau \frac{dy^\mu}{d\tau} \frac{dy^\nu}{d\tau} \frac{\delta^{(4)} (x-y(\tau))}{\sqrt{-g}}
\nonumber \\ &+ 8\pi \left(F^{\mu}_{~\rho} F^{\rho\nu} - \frac{1}{4} F_{\rho\sigma} F^{\rho\sigma} g^{\mu\nu} \right)\;.
\end{align}

In the following section, we make some simplification to solve the above equations \eqref{eq:fields} and compute the Proca fluxes emitted from EMRIs system.
Using the method developed in \cite{Spiers:2023cva}, one can expand the metric and vector field in power of $\eta$
\begin{eqnarray}
g^{\mu\nu} = g^{\mu\nu}_{(0)} + \eta h^{\mu\nu}_{(1)}  + \cdots \;, ~~ A^{\mu} = A^{\mu}_{(0)} +\eta A^{\mu}_{(1)}  + \cdots \;.
\end{eqnarray}
For zero order terms in mass ratio, the spacetime background is the Schwarzchild metric, and the vector field $A^{\mu}_{(0)}$ from $S_0$ is the constant, which can be set to zero without loss of generality.

The contributions of quantities $\mathcal{T}_{\rm Proca}^{\mu\nu}$ and $\alpha \mathcal{T}_{c}^{\mu\nu}$ coming from leading order in mass-ratio $\eta$ can be ignored, which exhibit only to higher orders in $\eta$ expansion.
Additionally, the actions $S_0$ and $S_c$ have dimensions, $[S_0]=(\rm mass)^2$ and
$[S_c]=(\rm mass)^{2-n}$; for an EMRIs system, $S_c$ describes the MBH spacetime, its dimension scale is mass $M$, so we can obtain the following relation
\begin{equation}
S_c \sim M^{-n} S_0.
\end{equation}
%Given that the parameter $\xi=\alpha/M^{n} = \eta^n \alpha/m^n$ and $\alpha/m^{n} \sim \mathcal{O}(1)$ due to the astrophysical constrains, the parameter satisfy $\xi\ll1$. As a result, $\alpha \mathcal{T}_{c}^{\mu\nu}$ can be ignored comparing to the Einstein tensor.
So one can continue to deduce that
\begin{equation}
\alpha \mathcal{T}_{c}^{\mu\nu} = -\frac{16\pi\alpha}{\sqrt{-g}} \frac{\delta S_c}{\delta g_{\mu\nu}} \sim -\frac{16\pi\alpha M^{-n}}{\sqrt{-g}} \frac{\delta S_0}{\delta g_{\mu\nu}},
\end{equation}
and
\begin{equation}
\alpha \mathcal{T}_{c}^{\mu\nu} \sim \xi G^{\mu\nu}\ll G^{\mu\nu}.
\end{equation}
On the other hand, using the similar method, we can get the approximate relation for vector field equation
\begin{align}
\frac{8\pi\alpha}{\sqrt{-g}} \frac{\delta S_c}{\delta A^\mu} &\sim \frac{8\pi\alpha M^{-n}}{\sqrt{-g}} \frac{\delta S_0}{\delta A^\mu}
\nonumber \\ &\sim \xi \left(\nabla_\rho F^{\rho\mu}-\mu_v^2 A^\mu\right) \ll \left(\nabla_\rho F^{\rho\mu}-\mu_v^2 A^\mu\right).
\end{align}
Therefore, we can conclude the field equations at first order in mass-ratio using the Dirac delta function~\cite{Zhu:2018tzi,Burton:2020wnj}
\begin{eqnarray}
G^{\mu\nu}[h^{\mu\nu}_{(1)}] &=& 8\pi m_p  \int d\tau \frac{dy^\mu}{d\tau} \frac{dy^\nu}{d\tau} \frac{\delta^{(4)} (x-y(\tau))}{\sqrt{-g}} ~\;, \label{eq:einstein} \nonumber \\
\nabla_{\rho} F^{\rho\mu}_{(1)}-\mu_v^2 A^{\mu}_{(1)}& =& 4\pi q m_p  \int d\tau \frac{dy^\mu}{d\tau} \frac{\delta^{(4)} (x-y(\tau))}{\sqrt{-g}}  ~\;,\label{eq:vector}
\end{eqnarray}
where $F^{\rho\mu}_{(1)}= \nabla^\rho A_{(1)}^\mu - \nabla^\mu A_{(1)}^\rho$ is the strength of vector field at first order perturbation.
The subscript $``(1)"$ for the vector and gravitational perturbations in the following section would be emitted for brevity.
The gravitational field equation in Eq. \eqref{eq:einstein} can return to the GR case, and
the Proca field equation in Eq. \eqref{eq:vector} has a source term whose magnitude is related to the vector charge carried by a secondary body.
As a result, the question of computing EMRIs emission can be simplified as the
handle of Einstein equation in GR and a Proca field equation in the Schwarzchild spacetime, the differences of EMRIs evolution arise from the the vector charge and mass.

The spacetime background of MBH can be given by the Schwarzchild metric
\begin{eqnarray}
ds^2&=&g_{\mu\nu}dx^\mu dx^\nu \nonumber\\
&=&-f(r) dt^2 + \frac{dr^2}{f(r)} +r^2 \left( d\theta^2 + \sin^2 \theta d\phi^2 \right)
\end{eqnarray}
with $f(r) = 1 - 2M/r$.
Due to the spherical symmetry of Schwarzchild spacetime, one can decompose the Proca equations as four ordinary differential equation (ODE) using the vector harmonics \cite{Rosa:2011my}, where the four-vector harmonics $Z^{(i)lm}_{\mu}$ satisfies
the following orthogonality condition
\begin{eqnarray}
\int \left(Z_{\mu}^{(i)lm}\right)^\ast \eta^{\mu \nu} Z_\nu^{(i')l'm'} d\Omega = \delta_{ii'} \delta_{ll'} \delta_{mm'}~,
\end{eqnarray}
where $\eta^{\mu\nu}=\text{diag}[1, f^2, 1/r^2, 1/(r^2\sin^2\theta)]$ and $d\Omega=\sin\theta d\theta d\phi$.
The vector harmonic functions are as follows
\begin{eqnarray} \label{vector-harmonics}
Z_{\mu}^{(1)lm} &=& \left[ 1, 0, 0, 0 \right] Y^{lm}, \\
Z_{\mu}^{(2)lm} &=& \left[ 0, f^{-1}, 0, 0 \right] Y^{lm}, \\
Z_{\mu}^{(3)lm} &=& \frac{r}{\sqrt{l(l+1)}} \left[ 0, 0, \partial_\theta, \partial_\phi  \right] Y^{lm}, \\
Z_{\mu}^{(4)lm} &=&  \frac{r}{\sqrt{l(l+1)}} \left[ 0, 0, \frac{1}{\sin\theta} \partial_\phi, - \sin\theta~\partial_\theta \right] Y^{lm}~,
\end{eqnarray}
where $Y^{lm} \equiv Y^{lm}(\theta, \phi)$ is the ordinary spherical harmonics.
Using the vector spherical harmonics, the vector potential $A_\mu$ and current density $J_\mu$ can be decomposed into the Fourier harmonic components
\begin{eqnarray}\label{ansatz}
\begin{split}
&A_{\mu}(t,r,\theta,\phi) =\frac{1}{r}\int_{-\infty}^{+\infty}d\omega~ \sum_{i=1}^{4} \sum_{lm} c_i \, u^{lm}_{(i)}(\omega,r) Z_\mu^{(i)lm}(\theta, \phi) e^{-i\omega t}~,\\
&J_{\mu}(t,r,\theta,\phi) = \frac{1}{r}\int_{-\infty}^{+\infty}d\omega~ \sum_{i=1}^{4} \sum_{lm} c_i \, s^{lm}_{(i)}(\omega,r) Z_\mu^{(i)lm}(\theta, \phi) e^{-i\omega t}~,
\end{split}
\end{eqnarray}
where $c_1 = c_2 = 1$, $c_3 = c_4 = [l(l+1)]^{-1/2}$.
Then the Proca field equation (\ref{eq:vector}) is recast as four second-order ODEs
\begin{widetext}
\begin{eqnarray}
\Op \afn_{(1)} + \left[ \frac{2}{r^2} \left(  \dot{\afn}_{(2)} - \afn^\prime_{(1)} \right) \right] &=&4\pi f s_{(1)} ,\label{eq-alp1} \\
\Op \afn_{(2)} + \frac{2}{r^2}\left[ \left(  \dot{\afn}_{(1)} - \afn^\prime_{(2)} \right) - f^2\left( \afn_{(2)} - \afn_{(3)} \right) \right]
&=&4\pi f s_{(2)},  \label{eq-alp2}\\
\Op \afn_{(3)} + \left[ \frac{2 f l (l+1)}{r^2} \afn_{(2)} \right]&= &4\pi f s_{(3)}, \label{eq-alp3}  \\
\Op \afn_{(4)} &=& 4\pi f s_{(4)}~,   \label{odd-parity}
\end{eqnarray}
\end{widetext}
where $s_{(1,2,3,4)}$ are four expansion coefficients of four-current for Proca field and the differential operator $\Op$ is
\beq
\Op \equiv \frac{\partial^2}{\partial r_\ast^2}+ \omega^2 - f \left[ \frac{l(l+1)}{r^2} + \mu_v^2 \right]~.
\eeq
where $\partial/\partial r_\ast = f(r)^{-1}$ is used to define the tortoise coordinate where the radial coordinate can be given by $r_\ast = r+2\ln\left(r/2M-1\right)$.
The Eq.~(\ref{odd-parity}), in fact, describes the odd-parity sector, which is independent on the other three equations (\ref{eq-alp1})--(\ref{eq-alp3}) describing the even-parity sector. With the Lorenz condition \cite{Rosa:2011my},
the Eq.~(\ref{eq-alp2}) can be reduced to
\begin{eqnarray}
\Op \afn_{(2)} - \frac{2 f}{r^2} \left(1 - \frac{3}{r} \right) \left( \afn_{(2)} - \afn_{(3)} \right) = 4\pi f s_{(2)}  \label{eq-alp2-alt} .
\end{eqnarray}
Finally, the four equations \eqref{eq-alp1}--\eqref{odd-parity} are simplified as one decoupled wave equation describing the odd-parity part and two coupled wave equations describing the even-parity mode.

\subsection{Orbital frequency and source term}
Because of the present of the spherical symmetry of Schwarzchild BH, we can  consider the time-like geodesics on the equatorial plane $(\theta=\pi/2)$ without loss of generality, which can be given by
\begin{equation}
\begin{split}
\frac{dt}{d\tau}&= E_p\left(1-\frac{2M}{r}\right)^{-1},\\
\frac{dr}{d\tau}&=V(r)^2 = E_p^2 - \left(1-\frac{2M}{r}\right) \left(1+\frac{J_p^2}{r^2}\right) ,\\
\frac{d\phi}{d\tau}&= \frac{J_p}{r^2}\;,
\end{split}\label{geo:eqs}
\end{equation}
where $V(r)$ is the effective radial potential, $E_p$ and $J_p$ is the orbital energy and angular momentum of secondary per unit mass $m_p$, respectively.
As the right hand of radial geodesic equation changes its sign when secondary approaches to the apsides of eccentric orbits, one appropriate method to describe the radial motion is the parameterized the geodesic equation by introducing a true anomaly $\chi$
\begin{equation}
r(\chi) = \frac{pM}{1+e\cos\chi},
\end{equation}
where $p$ is the dimensionless semi-latus rectum of orbit and $e$ is the eccentricity of orbit.
The radial geodesic equation satisfies the conditions at the apoapsis and
periapsis of eccentric orbits: $V(r_a)=V(r_p)=0$, where $r_p=pM/(1-e)$ is the apoapsis and $r_a=pM/(1+e)$ is the periapsis. Using two equations, we can obtain the analytical expression of constants of motion
\begin{eqnarray}
E_p^2 = \frac{(p-2)^2-4e^2}{p(p-3-e^2)}\;, \quad
J_p = \frac{pM}{\sqrt{p-3-e^2}}\;,\label{ELz}
\end{eqnarray}
then integrate the geodesic equations by substituting Eqs. \eqref{ELz} into Eqs.~\eqref{geo:eqs}
\begin{eqnarray}
t(\chi) &=& p^2M\sqrt{(p-2)^2-4e^2} \int_0^\chi (p-2-2e\cos\chi)^{-1} \nonumber\\ &\times& (1+e\cos\chi)^{-2} (p-6-2e\cos\chi)^{-1} d\chi\;, \\
\phi(\chi) & =& \sqrt{p}\int_0^\chi(p-6-2e\cos\chi)^{-1} d\chi\;.
\end{eqnarray}
The analytic expressions of two geodesic equations are obtained with the elliptic integrals in Ref. \cite{Wei:2025lva}. We can compute the radial and azimuthal orbital frequencies by defining the following equations
\begin{eqnarray}
\Omega_r = \frac{2\pi}{T_r} \;,\\
\Omega_\phi = \frac{\Delta\phi}{T_r} \;,
\end{eqnarray}
where $T_r = t(2\pi)$ is the period of radial motion and $\Delta\phi = \phi(2\pi)$ is the period of azimuthal motion for the point particle on the geodesic orbits.
The orbital frequency can be linearly combined with the radial and azimuthal frequency
\begin{equation}
\omega_{mn} = m\Omega_r + n \Omega_\phi\;,
\end{equation}
where $m, n$ are integers indices specifying the relative weights of the radial and azimuthal frequency components in the total orbital frequency.
The GW accumulated phase can be obtained by integrating observation time of LISA-like detectors
\begin{equation}
\Phi_{mn}  =2 \int_0^{t_{obs}}  \omega_{mn} (t') dt' \;,
\end{equation}
where $t_{obs}$ is the observational time. Similarly, we can define the radial
and azimuthal phases as follows
\begin{eqnarray}
\Phi_{r,\phi}  = 2\int_0^{t_{obs}}  \Omega_{r,\phi} (t') dt' \;.
\end{eqnarray}
In order to evaluate the difference of GW phase using two fundamental frequencies, we can define the radial and azimuthal dephasing via computing the equations
\begin{equation}
 \delta \Psi_{r,\phi} = \Phi_{r,\phi}^{\mu,q} - \Phi_{r,\phi}^{\rm GR}\;,
\end{equation}
where $\Phi_{r,\phi}^{\rm GR}$ denote to the GW phase using standard GR fluxes evolving inspiral and $\Phi_{r,\phi}^{\mu,q}$ is the GW phase computed with the Proca $(\mu\neq0,q\neq0)$ or massless  $(\mu=0,q\neq0)$ vector fluxes.

For the motion of secondary on the equatorial plane acting as the source term, the vector current density can be written down
\begin{equation}
J^\mu=q\frac{u^\mu}{u^t r_p^2 \sin\theta_p}\delta(r-r_p)\delta(\theta-\theta_p)\delta(\phi-\phi_p),
\end{equation}
where $u^\mu$ is four vector of secondary object
\begin{equation}
u^\mu = (E_p/f,0,0,J_p/r^2).
\end{equation}
The source terms for the Proca field equation in Eqs. \eqref{eq-alp2}-\eqref{odd-parity} are given by \cite{Zi:2024lmt}
\begin{equation}
\begin{split}
s_{(2)}&=0,\\
s_{(3)}&=-q\frac{\partial_\phi Y^{lm}(\pi/2,0)}{r_p^{3/2}}\delta(r-r_p)\delta(\omega-\omega_p),\\
s_{(4)}&=-q \frac{\partial_\theta Y^{lm}(\pi/2,0)}{r_p^{3/2}}\delta(r-r_p)\delta(\omega-\omega_p).
\end{split}
\end{equation}
The source terms for the gravitational equation have been widely studied in Refs. \cite{Cutler:1994pb,Martel:2003jj,Warburton:2011fk,Barack:2011ed,Akcay:2010dx}, which can be used to compute the gravitational fluxes with the Regge-Wheeler equations.

\subsection{Proca and gravitational fluxes}
To compute the massive vector radiation emitted by secondary objects endowed with Proca hair in EMRIs systems, we adopt a Green's function approach to obtain the nonhomogeneous solutions to Eqs.~\eqref{eq-alp1}--\eqref{odd-parity}. The procedure begins by solving the homogeneous equations for both even-parity and odd-parity modes. Then these homogeneous solutions are integrated with the corresponding source terms to construct the inhomogeneous solutions, enabling the evaluation of Proca fluxes.

The  homogeneous solutions for the odd-parity and even-parity equations can be obtained by imposing the ingoing and outgoing boundary conditions respectively; the detailed procedures are outlined in Refs.~\cite{Zhu:2018tzi,Zi:2024lmt}.
Within the adiabatic approximation, the inspiraling orbits of secondaries can be regarded to approximately follow the eccentric geodesics on the equatorial plane. The evolution of the orbital parameters is updated by the EMRIs fluxes, encompassing the Proca and gravitational radiation.
Given that the gravitational fluxes in the Schwarzchild spacetime have been studied maturely, these results are developed at the adiabatic order and post-adiabatic order (self-fore effect) \cite{Cutler:1994pb,Martel:2003jj,Poisson:2011nh,Warburton:2011fk,Barack:2011ed,Akcay:2010dx,Osburn:2015duj,Wardell:2021fyy,Barack:2019agd,Wei:2025lva}, the package of computing gravitational fluxes from Regge-Wheeler equation is also released at the website of  \texttt{Black Hole Perturbation Toolkit} \cite{BHPToolkit}, which is adopted to compute the gravitational fluxes.

In this section, we first focus on Proca fluxes from EMRIs system by computing the homogeneous solutions and source terms for the odd-parity and even-parity.
In the following section, we introduce the construct of general solutions using Green function.
For the odd-parity sector, the general solution can be given by
\begin{equation}
u_{(4)}=u_{(4)}^{H}\int_{r}^{\infty}d r_\ast~ \frac{4\pi f s_{(4)}}{W}u_{(4)}^{\infty}+u_{(4)}^{\infty} \int_{r_h}^{r}d r_\ast~ \frac{4\pi f s_{(4)}}{W}u_{(4)}^{H} \, ,
\end{equation}
with the wronskian
\begin{equation}
W=\frac{du_{(4)}^{\infty}}{d r_\ast}u_{(4)}^{H}- \frac{du_{(4)}^{H}}{d r_\ast}u_{(4)}^{\infty}\;,
\end{equation}
where $r_h=2M$ is the location of event horizon.
We can compute the general solutions in terms of the homogeneous solutions $u_{(4)}^{\infty,H}$ at infinity and near the horizon
\begin{equation}
\begin{split}
\lim_{r \to r_h}u_{(4)}&=e^{-i \omega r_\ast(r)}\int_{r_h}^{\infty}dr' ~\frac{4\pi f s_{(4)}}{W} u_{(4)}^{H}\\
&=\mathcal{A}^{4-}_{lm}e^{-i \omega r_\ast(r)}, \\
\lim_{r_\ast\to +\infty}u_{(4)}&= e^{i \sqrt{\omega^2-\mu_v^2} r_\ast(r)}r^{\frac{i\mu_v^2}{\sqrt{\omega^2-\mu_v^2}}}\int_{r_h}^{\infty}d r_\ast~\frac{4\pi f s_{(4)}}{W} u_{(4)}^{\infty} \\
&=\mathcal{A}^{4+}_{lm}e^{i \sqrt{\omega^2-\mu_v^2} r_\ast(r)}r^{\frac{i\mu_v^2}{\sqrt{\omega^2-\mu_v^2}}}.
\end{split}
\end{equation}
Using the expressions of source terms and changing integration variable,
one can get the functions $\mathcal{A}^{4-,+}_{lmn}$ integrating the parameter $\chi$ with the source term
\begin{equation}
\begin{split}
\mathcal{A}^{4-,+}_{lmn} = \int_{0}^{2\pi}d\chi' ~\frac{4\pi f(r) s_{(4)}(r)}{W(r)} u_{(4)}^{H,\infty}(r)e^{i\omega_{mn} t -im \phi(t)},
\end{split}
\end{equation}
where $t\equiv t(\chi)$, $\omega_{mn} = \omega_p$, $r\equiv r(\chi)$ and the superscripts $\pm$
denote to the up-going and in-gong homogeneous solutions.

For even-parity section, the ODEs can be recast into the differential equation matrix using the method described in Ref. \cite{Zhu:2018tzi,Burton:2020wnj}
\begin{align}
\label{eq:master}
\left( \frac{d^2}{dr_*^2} + \omega^2 + \left[\begin{array}{cc} \alpha_{lm} & \beta_{lm} \\ \gamma_{lm} & \sigma_{lm} \end{array} \right] \right) \left[\begin{array}{c} u_{(2)} \\ u_{(3)} \end{array} \right] &= \left[\begin{array}{c} S_{lm} \\ Z_{lm} \end{array} \right],
\end{align}
where the coefficients $\alpha_{lm}$, $\beta_{lm}$, $\gamma_{lm}$, $\sigma_{lm}$ are defined as follows
\begin{equation}
\begin{split}
\alpha_{lm}&=-f \left[ \frac{l(l+1)}{r^2} + \mu_v^2 \right]-\frac{2 f}{r^2} \left(1 - \frac{3}{r} \right),\\
\beta_{lm}&=\frac{2 f}{r^2} \left(1 - \frac{3}{r} \right),\\
\gamma_{lm}&=\frac{2 f l (l+1)}{r^2},\\
\sigma_{lm}&=-f \left[ \frac{l(l+1)}{r^2} + \mu_v^2 \right].
\end{split}
\end{equation}

Given that the asymptotic behavior of the quantities \(\alpha_{lm}\), \(\beta_{lm}\), and \(\gamma_{lm}\) at infinity and near the horizon approach to zero, we do not present their explicit forms here. The homogeneous solutions $u_{(2,3,4)}^{\infty,H}$ at infinity and near the horizon can be imposed the out-going and in-going asymptotic behaviors, which have been systematically argued in Ref.~\cite{Zi:2024lmt}, interested readers can refer to Refs. \cite{Zhu:2018tzi,Burton:2020wnj,Zi:2024lmt} for further details.

The source term in the Eq. \eqref{eq:master} can be expressed as following form using the Dirac delta functions
\begin{eqnarray}
\label{eq:source}
\left[\begin{array}{c} S_{lm}(r) \\ Z_{lm}(r) \end{array} \right] &=& \left[\begin{array}{c} B_{lm} \\ D_{lm} \end{array} \right]\delta(r-r_0) \nonumber \\ &+& \left[\begin{array}{c} F_{lm} \\ H_{lm} \end{array} \right] \frac{d}{dr}\delta(r-r_0) ,
\end{eqnarray}
where $B_{lm}$, $D_{lm}$, $F_{lm}$, and $H_{lm}$ can be given by the orbital elements of the point particle.
In terms of the source terms and homogeneous solutions, the inhomogeneous solution can be expressed as a piecewise function as following \cite{Zhu:2018tzi}
\begin{align}
\label{eq:inhomo}
&\left[\begin{array}{c} u_{(2)} \\ u_{(3)} \end{array} \right] = \left( \mathcal{A}^{2+}_{lm} \left[ \begin{array}{c} u_{(2)}^{0+} \\ u_{(3)}^{0+} \end{array} \right] + \mathcal{A}^{3+}_{lm} \left[ \begin{array}{c} u_{(2)}^{1+} \\ u_{(3)}^{1+} \end{array} \right]  \right) \Theta(r-r_0) \notag
\\&\qquad\;\;\; +\left( \mathcal{A}^{2-}_{lm} \left[ \begin{array}{c} u_{(2)}^{0-} \\ u_{(3)}^{0-} \end{array} \right] + \mathcal{A}^{3-}_{lm} \left[ \begin{array}{c} u_{(2)}^{1-} \\ u_{(3)}^{1-} \end{array} \right]  \right) \Theta(r_0-r) ,
\end{align}
where $\Theta$ is the Heaviside step function and the quantities $\mathcal{A}^{2,3\pm}_{lm} $
\begin{eqnarray}
\mathcal{A}^{2,3-}_{lmn}  &=&\int_{r_h}^{r}dr'~ \frac{4\pi f s_{(2,3)}}{W} u_{(2,3)}^{H} \;, \\
\mathcal{A}^{2,3+}_{lmn} &=&\int_{r}^{\infty}dr'~\frac{4\pi f s_{(2,3)}}{W} u_{(2,3)}^{\infty} \;.
\end{eqnarray}
The four above equations can transformed as the integrals of parameter $\chi$:
\begin{eqnarray}
\mathcal{A}^{2,3-}_{lmn}  &=&\int_{0}^{2\pi}d\chi'~ \frac{4\pi f(r) s_{(2,3)}(r)}{W(r)} u_{(2,3)}^{H}(r) \;, \\
\mathcal{A}^{2,3+}_{lmn} &=& \int_{0}^{2\pi}d\chi'~\frac{4\pi f(r) s_{(2)}(r) }{W(r)} u_{(2,3)}^{\infty}(r) \;
\end{eqnarray}
with a function $r\equiv r(\chi)$,  we replace the superscripts $(H,\infty)$ with symbols $\pm$ to to simplify the notation.

Using these inhomogeneous solutions for the even-parity and odd-parity modes, we can compute Proca fluxes from EMRIs, the following section introduces the relevant expression.
The Proca fluxes can be derived from the Isaacson's stress-energy tensor \cite{Isaacson:1968zza}
\begin{equation}\label{Isaacson:tensor}
4\pi T_{\mu\nu}= g^{\alpha\beta}F_{\mu \alpha}F_{\nu \beta}+\mu_v^2 A_{\mu}A_{\nu}-g_{\mu\nu}\mathcal{L}.
\end{equation}
Starting from the formula \eqref{Isaacson:tensor}, one can define the vector flux thought a surface $\Sigma^\pm$ over the time $dt$
\begin{equation}
dE^{\pm} = \mp \int T^\mu_{~\nu}~ \xi^\nu_{(t)} d\Sigma{_\mu},
\end{equation}
where $d\Sigma_\mu$ is the outward oriented surface element on the surface $\Sigma$, the integral is computed over the 2-spheres at the constant $r$.
The negative sign denote to the energy flux at infinity, and the positive sign is the energy flux near the horizon.
The Schwarzschild black hole owns the static and axially symmetries,
which corresponds to two Killing vectors, $\xi^\alpha_{(t)}$ and $\xi^\alpha_{(\phi)}$, the Proca flux expression can be simplified as
\begin{align} \label{dEdt:Formular}
\frac{dE^{ \mathcal{P}}}{dt} =
\begin{cases}
\int T_{tr}~ r^2 f  d\Omega \quad~  r\rightarrow 2M  \;,  \\
- \int T_{tr}~ r^2 f  d\Omega \quad ~  r\rightarrow \infty \;,
\end{cases}
\end{align}
where $d\Omega =\sin\theta d\theta d\phi$.
After some simplification of the expression in Eq.~\eqref{dEdt:Formular}, the energy flux formulas at infinity for the even and odd sectors are reduced as follows
\begin{equation}\label{eq:Edot:P:I:odd}
\left\langle \frac{d E ^{ \mathcal{P}}}{dt}\right\rangle_{\rm odd}^{\infty}=\sum_{l=1}^\infty \sum_{m=1}^l \sum_{n=1}^{\infty}
\frac{\omega\sqrt{\omega^2-\mu_v^2}|\mathcal{A}^{4+}_{lmn}|^2}{2\pi l(l+1)}\,,
\end{equation}
\begin{eqnarray}\label{eq:Edot:P:I:even}
\left\langle \frac{d E ^{ \mathcal{P}}  }{dt}\right\rangle_{\rm even}^{\infty}&=&\sum_{l=1}^\infty \sum_{m=1}^l \sum_{n=1}^{\infty}
\frac{\omega\sqrt{\omega^2-\mu_v^2}|\mathcal{A}^{3+}_{lmn}|^2}{2\pi l(l+1)}\\ \nonumber &+&\frac{\mu^2}{2\pi}\frac{\sqrt{\omega^2-\mu_v^2}}{\omega}|\mathcal{A}^{2+}_{lmn}|^2\;,
\end{eqnarray}
and the energy flux formulas near the horizon can be given by
\begin{equation}\label{eq:Edot:P:H:odd}
\left\langle \frac{d E ^{ \mathcal{P}}}{dt}\right\rangle_{\rm odd}^{H}=\sum_{l=1}^\infty \sum_{m=1}^l \sum_{n=1}^{\infty} \frac{\omega^2|\mathcal{A}^{4-}_{lmn}|^2}{2\pi l(l+1)}\,,
\end{equation}
\begin{eqnarray}\label{eq:Edot:P:H:even}
\left\langle \frac{d E^{ \mathcal{P}}}{dt}\right\rangle_{\rm even}^{H} &=&
\sum_{l=1}^\infty \sum_{m=1}^l \sum_{n=1}^{\infty} \frac{\omega^2|\mathcal{A}^{3-}_{lmn}|^2}{2\pi l(l+1)}
\\ \nonumber &+&
\frac{l(l+1)+4\mu_v^2}{8\pi}|\mathcal{A}^{2-}_{lmn}|^2
\\ \nonumber &-&\frac{i\omega}{4\pi}\left(\mathcal{A}^{2-}_{lmn} \mathcal{A}^{3-\ast}_{lmn}-\mathcal{A}^{3-}_{lmn} \mathcal{A}^{2-\ast}_{lm}\right)\,.
\end{eqnarray}
The angular momentum fluxes can be obtained with the following relation
\begin{eqnarray}\label{J:dots:H:infty}
\left\langle \frac{d J^{ \mathcal{P}}}{dt}\right\rangle_{\rm even, odd}^{H,\infty}
= \frac{m}{\omega_{mn}} \left\langle \frac{d E^{ \mathcal{P}}}{dt}\right\rangle_{\rm even, odd}^{H,\infty} \; .
\end{eqnarray}
The full details of the derivation for the flux formulas have been shown in the Appendix A of Ref.~\cite{Zi:2024lmt}.
Therefore, the total fluxes from EMRIs system contain the contributaion of the gravitational emission and the Proca emission
\begin{eqnarray}
\dot{E} &=& \dot{E}^\mathcal{G} +\dot{E}^\mathcal{P}\;, \label{Edot:tot} \\
\dot{J} &=& \dot{J}^\mathcal{G} +\dot{J}^\mathcal{P}\;,\label{Jdot:tot}
\end{eqnarray}
where
\begin{eqnarray}
\dot{E}^{\mathcal{P}} &=& \dot{E}^{\mathcal{P},H}_{\rm odd} + \dot{E}^{\mathcal{P},H}_{\rm even}
+ \dot{E}^{\mathcal{P},\infty}_{\rm odd} + \dot{E}^{\mathcal{P},\infty}_{\rm even}\;, \label{Edot:tot:G:P} \\
\dot{J}^{\mathcal{P}} &=& \dot{J}^{\mathcal{P},H}_{\rm odd} + \dot{E}^{\mathcal{P},H}_{\rm even}
+ \dot{J}^{\mathcal{P},\infty}_{\rm odd} + \dot{J}^{\mathcal{P},\infty}_{\rm even} \;, \label{Jdot:tot:G:P}
\end{eqnarray}
here the quantities with a superscript dot denote to the fluxes
\begin{eqnarray}
\dot{E} &=& \left\langle \frac{dE}{dt} \right\rangle\;, \label{Edot:tot:symbol} \\
\dot{J} &=& \left\langle \frac{dJ}{dt} \right\rangle\;, \label{Jdot:tot:symbol} \\
\dot{E}^{\mathcal{G},\mathcal{P}}_{\rm even, odd} &=& \left\langle \frac{dE^{\mathcal{G},\mathcal{P}}}{dt} \right\rangle _{\rm even, odd}^{H,\infty}, \label{Edot:tot:odd:even:symbol} \\
\dot{J}^{\mathcal{G},\mathcal{P}}_{\rm even, odd} &=& \left\langle \frac{dJ^{\mathcal{G},\mathcal{P}}}{dt} \right\rangle _{\rm even, odd}^{H,\infty}\;.
\label{Jdot:tot:odd:even:symbol}
\end{eqnarray}
For the gravitational radiation, we can generate the gravitational fluxes $(\dot{E}^{\mathcal{G}}, \dot{J}^{\mathcal{G}})$ using Regge-Wheeler {\bf Mathemtica} package. Moreover, for the Proca radiation, we can compute the Proca fluxes $(\dot{E}^{\mathcal{P}}, \dot{J}^{\mathcal{P}})$ with Eqs.~\eqref{eq:Edot:P:I:odd}--\eqref{eq:Edot:P:H:even}.
Note that the summation of total fluxes $\dot{E}$ and $\dot{J}$ are truncated
by running the indexes $(l,m,n)$ in Eqs.~\eqref{eq:Edot:P:I:odd}--\eqref{J:dots:H:infty}, where we chose the maximum of index as $l_{\text{max}}=6,~ n_{\text{max}} = 10$ to reach the absolute difference $\leq 10^{-4}$ comparing to the higher modes.
We list the energy fluxes (in units of $m^2_p/M^2$) from the gravitational and vector emissions for the modes $n_{\text{max}}=10,~l_{\text{max}}=11$ in Table \ref{Tab:Fluxvalues}, and the fluxes is summed running the index from $l_{\rm min} \in \{1, 2\}$ to $l_{\text{max}}=10$, where $l_{\rm min}=1$ corresponds to the minimum index of  vector fluxes summation
and $l_{\rm min}=2$ corresponds to  the minimum index of the gravitational fluxes summation.
One can observe that the contribution of higher mode $n=11$ is less than $10^{-4}$ for the eccentric EMRIs orbits $(e=0.18)$, the truncation error can also be a calibration of computing fluxes for the higher mode.
On the other hand, to examine the validity of our code, we also compute the gravitational flux setting same orbital parameters as Refs.~\cite{Cutler:1994pb,Zhang:2022rfr}, which is almost consistent with the previous result.

\begin{table*}[htbp!]
\centering
\begin{tabular}{ccccccccccc}
\hline
\hline
$\mu$ & $q$ &  $p$ & $e$ & $\dot{E}^{\mathcal{P},n_{\text{max}}=10}_\infty$
& $\dot{E}^{\mathcal{P},n_{\text{max}}=11}_\infty$  & Error
& $\dot{E}^{\mathcal{G},n_{\text{max}}=10}_\infty$
& $\dot{E}^{\mathcal{G},n_{\text{max}}=11}_\infty$
& Error. \\
\hline
0.0 &0.0 &7.5 &0.0  &$-$ &  $-$ & $-$  &$4.50436\times10^{-4}$ &  $4.50423\times10^{-4}$ & $<10^{-3} $ \\
0.0 &0.0 &12 &0.0 &$-$ &$-$  & $-$ &$4.00196\times10^{-5}$ &$4.00191\times10^{-5}$  & $<10^{-4}$ \\
\hline
0.0 &0.0 &7.5 &0.18 &$-$ &  $-$ & $-$  &$\color{red} 3.16676\times10^{-4}$ &  $3.16678\times10^{-4}$ & $<10^{-4} $ \\
0.0 &0.0 &12 &0.18 &$-$ &$-$  & $-$ &$2.92711\times10^{-5}$ &$2.92708\times10^{-5}$  & $<10^{-4} $\\
\hline
0.0 &0.1 &7.5  &0.18 &$1.02329\times10^{-4}$ &$2.57258\times10^{-4}$  & $<10^{-4} $ &$\checkmark $ &$\checkmark $  & $\checkmark $\\
0.0 &0.1 &12 &0.18 &$3.59615\times10^{-5}$   &$3.59608\times10^{-5}$  & $<10^{-4}$ &$\checkmark $ &$\checkmark $  & $\checkmark  $\\
\hline
0.01 &0.1 &7.5  &0.0 &$4.57231\times10^{-4}$ &$4.57258\times10^{-4}$  & $\leq10^{-4} $ &$\checkmark $ &$\checkmark $  & $\checkmark $\\
0.01 &0.1 &12 &0.0 &$3.00621\times10^{-5}$ &$3.00618\times10^{-5}$  & $<10^{-4}$ &$\checkmark $ &$\checkmark $  & $\checkmark  $\\
\hline
0.01 &0.1 &7.5  &0.18 &$1.02262\times10^{-4}$ &$1.02258\times10^{-4}$  & $<10^{-4} $ &$\checkmark $ &$\checkmark $  & $\checkmark $\\
0.01 &0.1 &12 &0.18 &$2.84867\times10^{-5}$  &$2.84862\times10^{-5}$  & $<10^{-4}$ &$\checkmark $ &$\checkmark $  & $\checkmark  $\\
\hline
\hline
\end{tabular}
\caption{The Proca  and gravitational fluxes (in units of $m_p^2/M^2$) emitted from the EMRIs system for the orbital geometric parameters $p\in\{7.5,12\}$ and $e\in\{0.0,0.18\}$ are listed, which consider the different vector fields with a mass $\mu \in \{0,0.01\}$ and charge $q\in \{0,0.1\}$.
The seventh column and tenth column are the relative difference of the vector flux for $n_{\text{max}}=10$ and $n_{\text{max}} = 11$, which correspond to the Proca and gravitational fluxes at the infinity.
Note that the parameters $\mu=0$ and $\mu\neq0$ denote the cases of the massless vector field and the Proca field.
Under the same setting of parameters, the symbols of transverse line in the table denote the absent of vector radiation emission in EMRIs systems, and the symbols of check mark
represent for the gravitational fluxes keeps same with the above listing the gravitational fluxes.
The gravitational flux of red font ``$3.16676\times10^{-4}$" is also
same almost with the result in Ref. \cite{Cutler:1994pb,Zhang:2022rfr}.
}\label{Tab:Fluxvalues}
\end{table*}

\subsection{Adiabatic evolution and waveform analysis}
In this subsection, we first briefly review the procedure of adiabatic evolution, where the orbital parameters are updated with the fluxes computed by Eqs. \eqref{Edot:tot}--\eqref{Jdot:tot}.
Specifically, the dissipation of the orbital energy and angular momentum due to the gravitational and scalar emission derives the evolution of EMRIs orbital parameters, in which the inspiraling trajectories can be regarded as a sequence of geodesics to the plunge by the primary.
In the adiabatic approximation, the varying ratio of orbital integrals $\dot{\mathcal{C}}\in\{E_{\rm orb},J_{\rm orb}\} $ can be determined with the balance law
\begin{equation}
\dot{\mathcal{C}} = -\dot{\mathcal{C}}_{\rm GW} \;,
\end{equation}
where $\dot{\mathcal{C}}_{\rm GW} \in \{\dot{E}, \dot{J}\}$ is given with Eq.~\eqref{Edot:tot}--\eqref{Jdot:tot}. The computation of two kinds of fluxes can be supplemented in the \texttt{FastEMRIWaveform} (\texttt{FEW}) model~\cite{Chua:2017ujo,Chua:2020stf,Katz:2021yft} to generate the EMRIs waveform fast and assess the effect of Proca emission.

In the following section, we describe the interpolation scheme used to generate fast fluxes for the orbital parameters \((p, e)\) using the relativistic flux data from Eqs.~\eqref{Edot:tot}--\eqref{Jdot:tot}. We first compute the Proca fluxes for both the even- and odd-parity sectors, together with the gravitational fluxes, on a rectangular grid of sampling points in the \((p, e)\) plane. The rectangular grid contains several hundred points. For a given Proca mass, we construct a two-dimensional uniform grid consisting of 35 samples in the eccentricity direction, spanning \(e \in [0, 0.7]\), and 50 samples in the semi-latus rectum direction defined by
\begin{equation}
p_i = p_{\rm min} + 4(e^{i\Delta u} - 1), \quad 0 \leq i \leq 49,
\end{equation}
where \(\Delta u = 0.02\) and \(p_{\rm min} = p_{\rm sp} + 0.02\), with \(p_{\rm sp} = 6M + 2e\) denoting the separatrix location. To maintain numerical stability, all sampling points are placed at least \(0.02\) away from the separatrix. The upper bound is set at \(p_0 = 12\), allowing reliable interpolation of the fluxes within \(p \in [p_{\rm min}, p_0]\) for fixed eccentricity. The numerical
scheme can ensure to rapidly generate the fluxes using
the interpolation method, which is used to adiabatically
evolve the orbital parameters in \texttt{FEW} model.

\begin{figure*}[htb!]
\centering
\includegraphics[width=3.17in, height=2.3in]{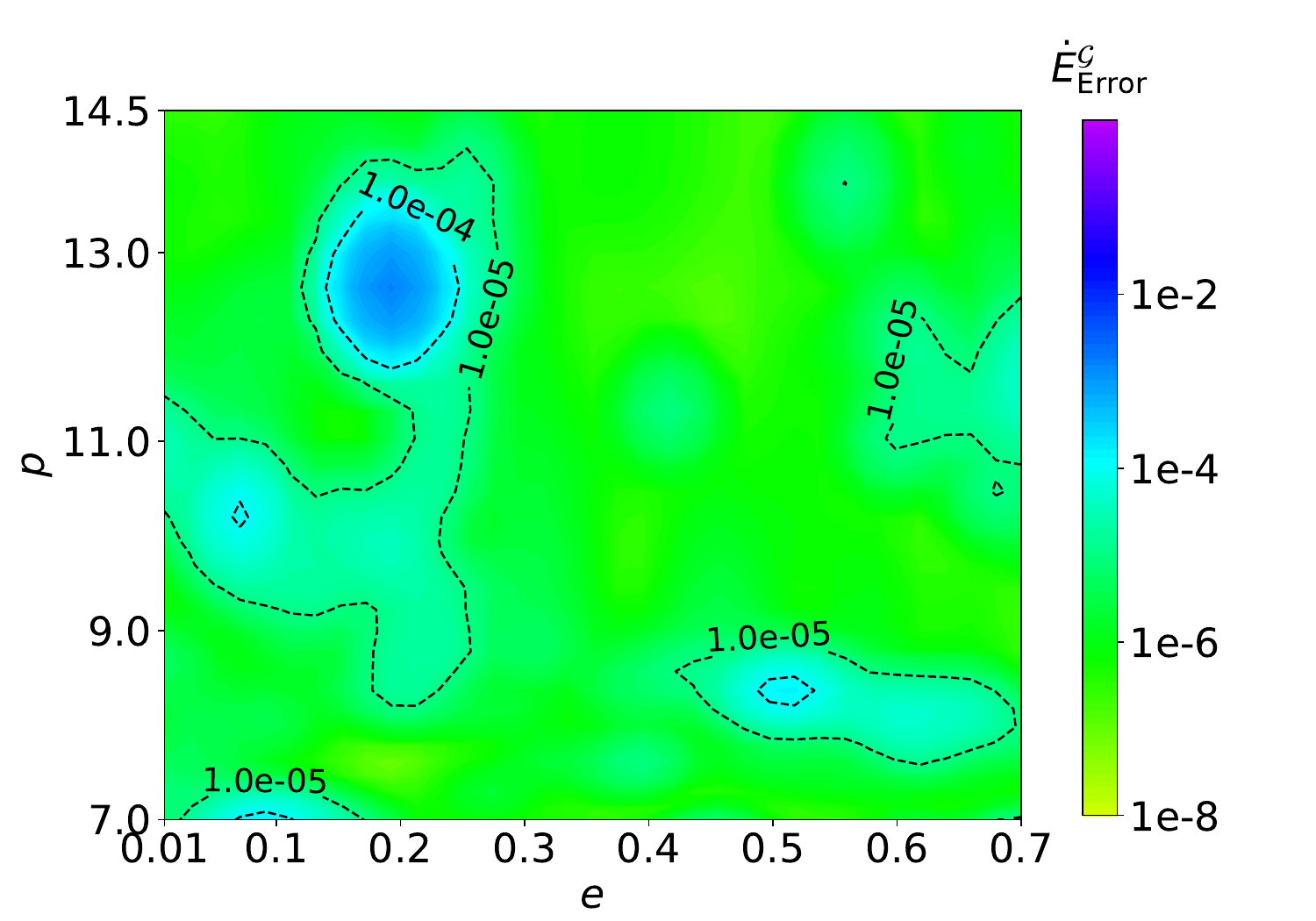}
\includegraphics[width=3.17in, height=2.3in]{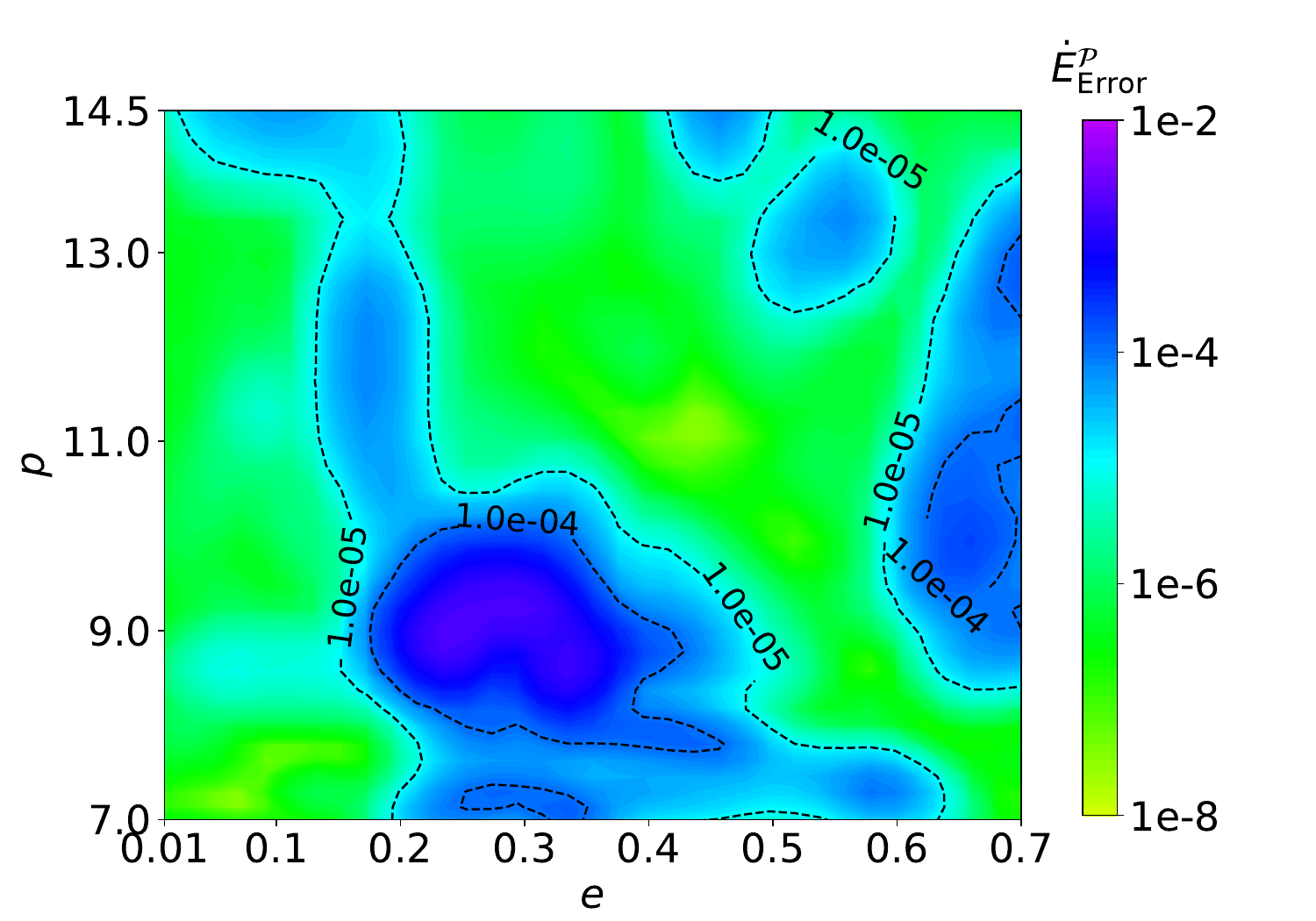}
\includegraphics[width=3.17in, height=2.3in]{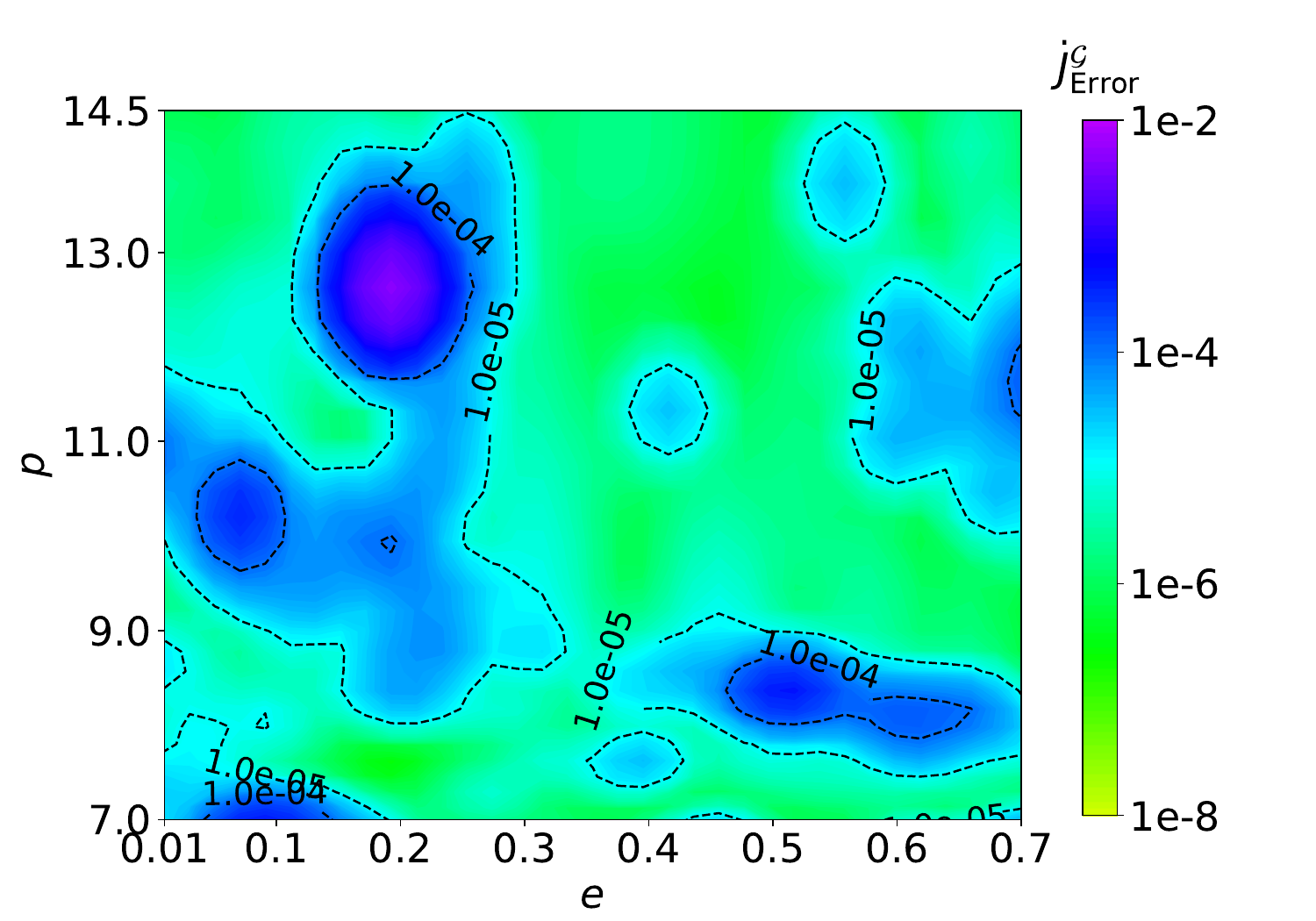}
\includegraphics[width=3.17in, height=2.3in]{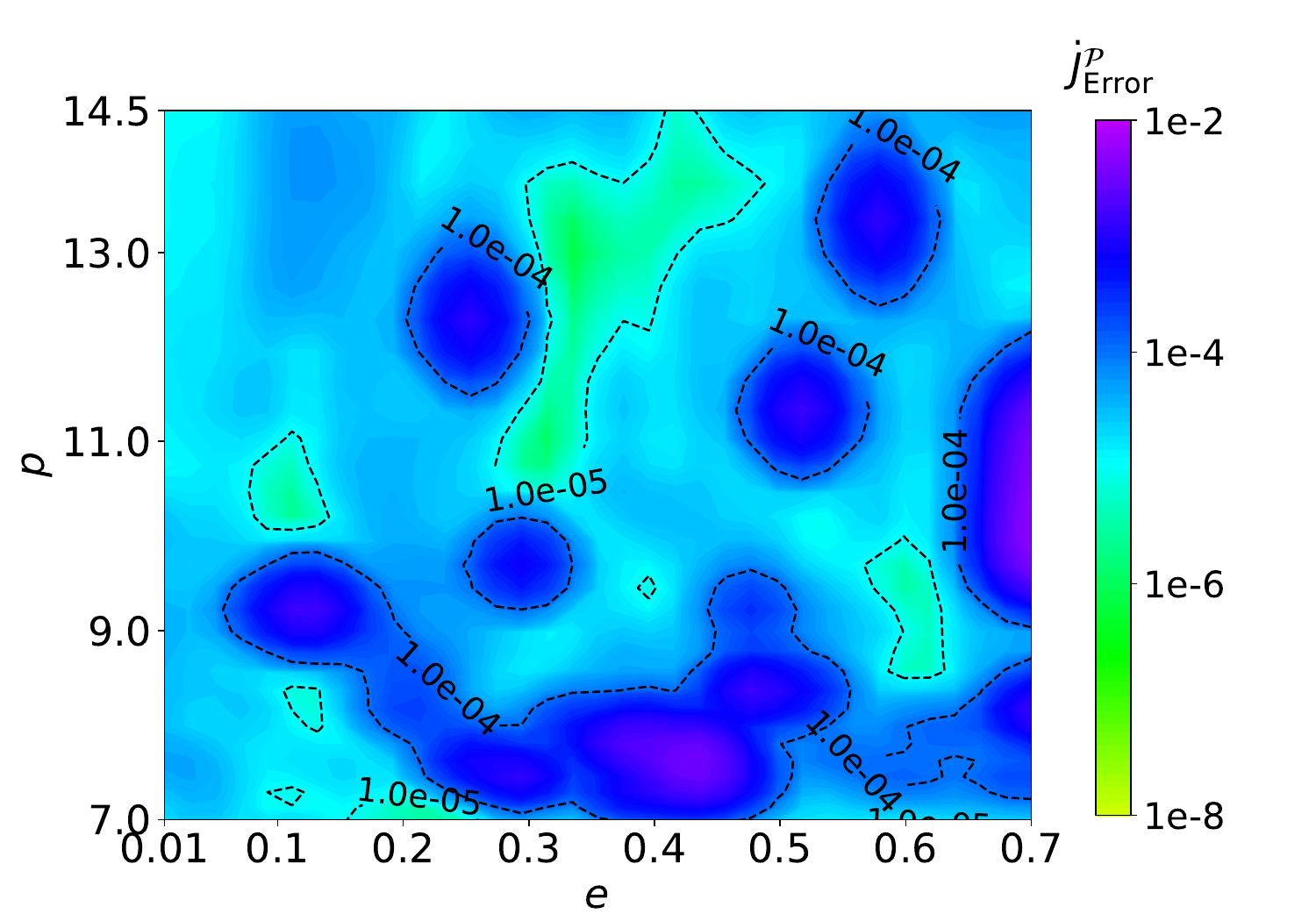}
\caption{Fluxes errors on the retangular grid between the interpolation method and the Proca or gravitational flux obtained over summing index $(l,m,n)$, as the functions of two orbital parameters $(p,e)$ are plotted, assuming the Proca mass $\mu=0.001$. The symbol $\dot{E}^{\mathcal{G}}_{\rm Error}$ is the error of the gravitational energy fluxes computed with two methods, $\dot{J}^{\mathcal{G}}_{\rm Error}$ is  the error for  the gravitational angular momentum fluxes, and the quantity with a subscript $\mathcal{P}$ denotes to the Proca flux error.
These energy fluxes have an unit of mass-ratio $m^2_p/M^2$ and the angular momentum fluxes possess an unit of $m^2_p/M$. } \label{Fig:fluxes:error}
\end{figure*}

To evaluate the error of computing fluxes using the interpolation method, Fig.~\ref{Fig:fluxes:error} shows the relative differences in the energy and angular momentum fluxes between between the relativistic results and those obtained by interpolation over the $(p,e)$ grid.
The relative differences of fluxes are the function of orbital parameters $(p,e)$ that constitutes the rectangular grids, the dashed lines represent for the contour lines of two fluxes differences. One can see that the interpolation errors of fluxes almost lie at $\sim 10^{-5}$, however, the fluxes errors is about $10^{-3}$ for some very few EMRIs systems.
The error distributions, in some ways, manifest our interpolation method can serve as a temporary plan in the orbital evolution and waveform generation of EMRIs.

After computing the energy and angular momentum fluxes on a two-dimensional grid in $(p,e)$ plane, we can model the inspiraling trajectory at the adiabatic order in \texttt{FEW} model, allowing to generate the waveform and implement the low-frequency approximation of LISA response.
In the framework of weaker field approxiamtion,
the waveform quadrupolar formula based on the pioneering work by Peter and Matthews includes the effects of pericenter and Lense-Thirring precession \cite{Peters:1963ux}. The strain of amplitudes responded by LISA can be approximately summed via some harmonics
\begin{equation}\label{antenna}
h_{\textup{I,II}} = \sum_n \frac{\sqrt{3}}{2} \Big[F^+_{\textup{I,II}} (t)A^+_n(t) + F^\times_{\textup{I,II}} (t)A^\times_n(t) \Big]\;,
\end{equation}
where the quantities $F^+_{\textup{I,II}}$ are the antenna pattern functions,
their full forms can refer to Ref.~\cite{Cutler:1997ta}.
The amplitudes $A^+_n(t)$ can be given by
\begin{align}\label{amplitude}
A_n^+ = &-\Big[1+(\hat{L}\cdot\hat{n})^2\Big]\Big[a_n \cos2\gamma -b_n\sin2\gamma\Big] +c_n\Big[1-(\hat{L}\cdot\hat{n})^2\Big], \nonumber\\
A_n^\times =& 2(\hat{L}\cdot\hat{n})\Big[b_n\cos2\gamma+a_n\sin2\gamma\Big],
\end{align}
where $\gamma$ is the angle directing the orbital pericenter and the coefficients are written as
\begin{equation}
\begin{aligned}
a_n =~ &-n \mathcal{A} \Big[J_{n-2}(ne)-2eJ_{n-1}(ne)+\frac{2}{n}J_n(ne)
\\ \nonumber &+2J_{n+1}(ne) -J_{n+2}(ne)\Big]\cos(n\Phi_r), \\
b_n =~ &-n \mathcal{A}(1-e^2)^{1/2}\Big[J_{n-2}(ne)-2J_{n}(ne)+J_{n+2}(ne)\Big]
\\ \nonumber &\times\sin(n\Phi_r), \\
c_n =~& 2\mathcal{A}J_n(ne)\cos(n\Phi_r), \\
\mathcal{A} = ~& (M \omega_\phi )^{2/3} m_p/D_L\;,
\end{aligned}
\end{equation}
with the $J_n$ is Bessel functions of the first kind.
In our case of EMRIs orbits on equatorial plane,  the angle $\gamma = \Phi_\phi - \Phi_r$ denotes the direction of eccentric orbit pericenter, and
the quantity $\hat{L}\cdot\hat{n}$ can be obtained in terms of four angles $(\theta_S,\phi_S,\theta_L, \phi_L)$ that determines the direction of source and orbital angular momentum \cite{Barack:2003fp} and $D_L$ is the distance from astrophysical source to the GW detector.

Note that the choice of waveform template do not have a significant influence on the measurement precision of intrinsic parameters $\Theta_i = (M,m_p, p_0, e_0, \Phi_{r,0}, \Phi_{\phi,0}, q, \mu)$ \cite{Katz:2021yft}, so the hybrid waveform model incorporating the relativistic trajectory and quadrupole waveform formula can be used to estimate the effect of lighter Proca field on EMRIs signal.
A preliminary assessment of the distinguishability of the Proca field mass can be executed by computing the mismatch between two EMRIs signals with and
without the correction of the Proca mass. The formula can be obtained by defining the overlap of two waveforms
\begin{align}\label{overlap}
\mathcal{M} = 1-\mathcal{O}(h_a, h_b) = 1- \frac{(h_{a}\vert h_{b})}{\sqrt{(h_{a}\vert h_{a})(h_{b}\vert h_{b})}},
\end{align}
where the inner product is defined by \cite{Cutler:1994ys}
\begin{equation}\label{overlap2}
(h_a|h_b) = 2 \int^{f_{\rm high}}_{f_{\rm low}} \frac{h_a^*(f)h_b(f)+h_a(f)h_b^*(f)}{S_n(f)}df \;.
\end{equation}
The function $S_n(f)$ is the frequency dependent noise-weighted sensitivity curve of the space-borne detector, such as LISA \cite{LISA:2017pwj}.
The range of integration is $f\in[f_{\textrm{low}}, f_{\textrm{high}}] $,
where $f_{\textrm{low}} = 0.1 \rm mHz$ and $f_{\textrm{high}}$ is the orbital frequency near the LSO ($f_{\rm LSO}$) or the orbital frequency after one year of evolution.
If the mismatch equals to zero, one can claim that two EMRIs signals can not be distinguished by the detectors.
An empirical rule to discern two GW signals  is
the minimum of mismatch is $\mathcal{M}_{c} \sim \frac{1}{2\rho^{2}}$ distinguished by the LISA-like detectors, where $\rho$ is the signal-to-noise ratio (SNR). For the moderate EMRIs event detected by LISA, the SNR can be take to $\rho = 20$ \cite{Babak:2017tow,Fan:2020zhy}.
Thus, the threshold value of the mismatch distinguished by LISA is $\mathcal{M}_{c} \sim 0.001$, which can be treated as the temporary criterion for recognizing two GW signals in the following analysis of EMRIs waveforms.

\begin{figure*}[htb!]
\centering
\includegraphics[width=3.17in, height=2.3in]{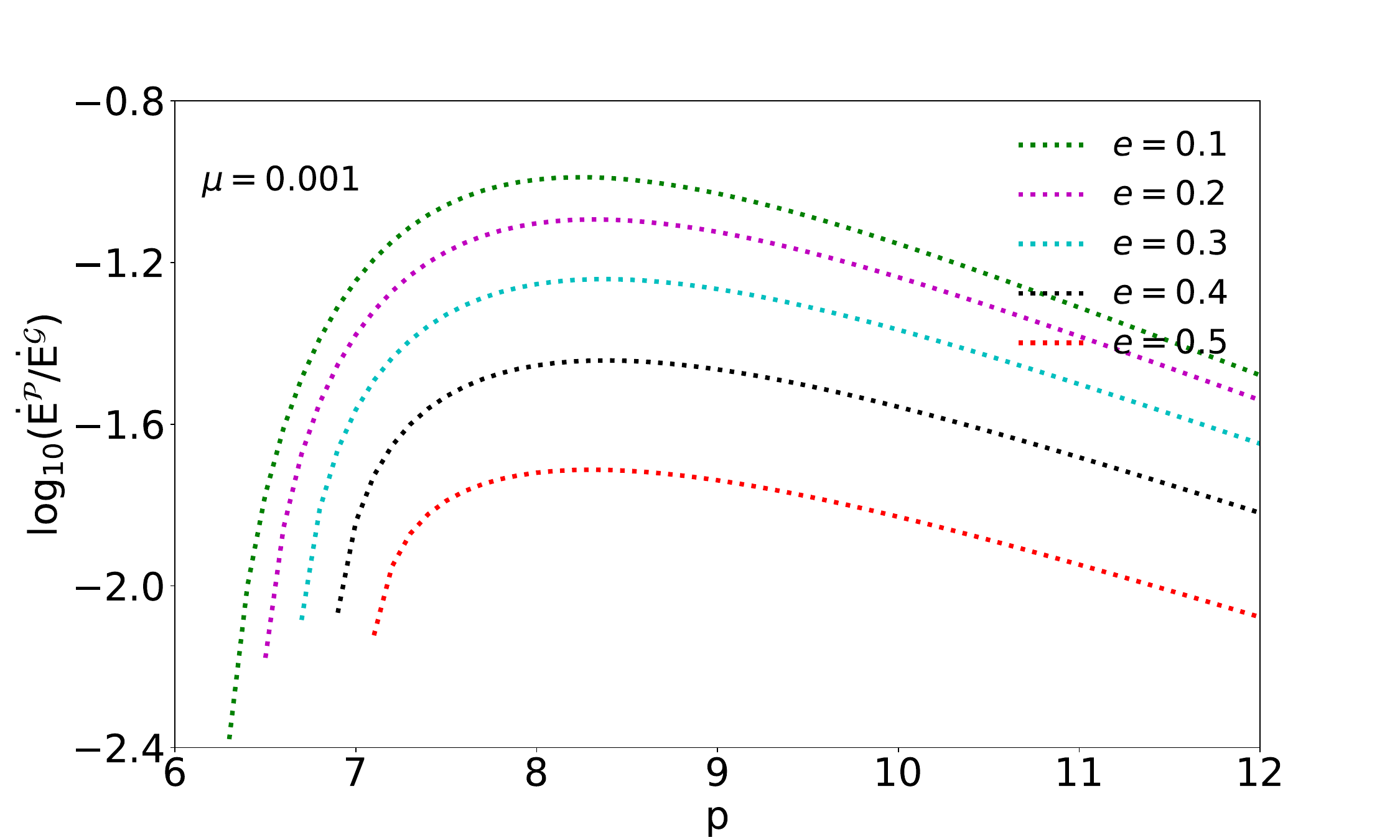}
\includegraphics[width=3.17in, height=2.3in]{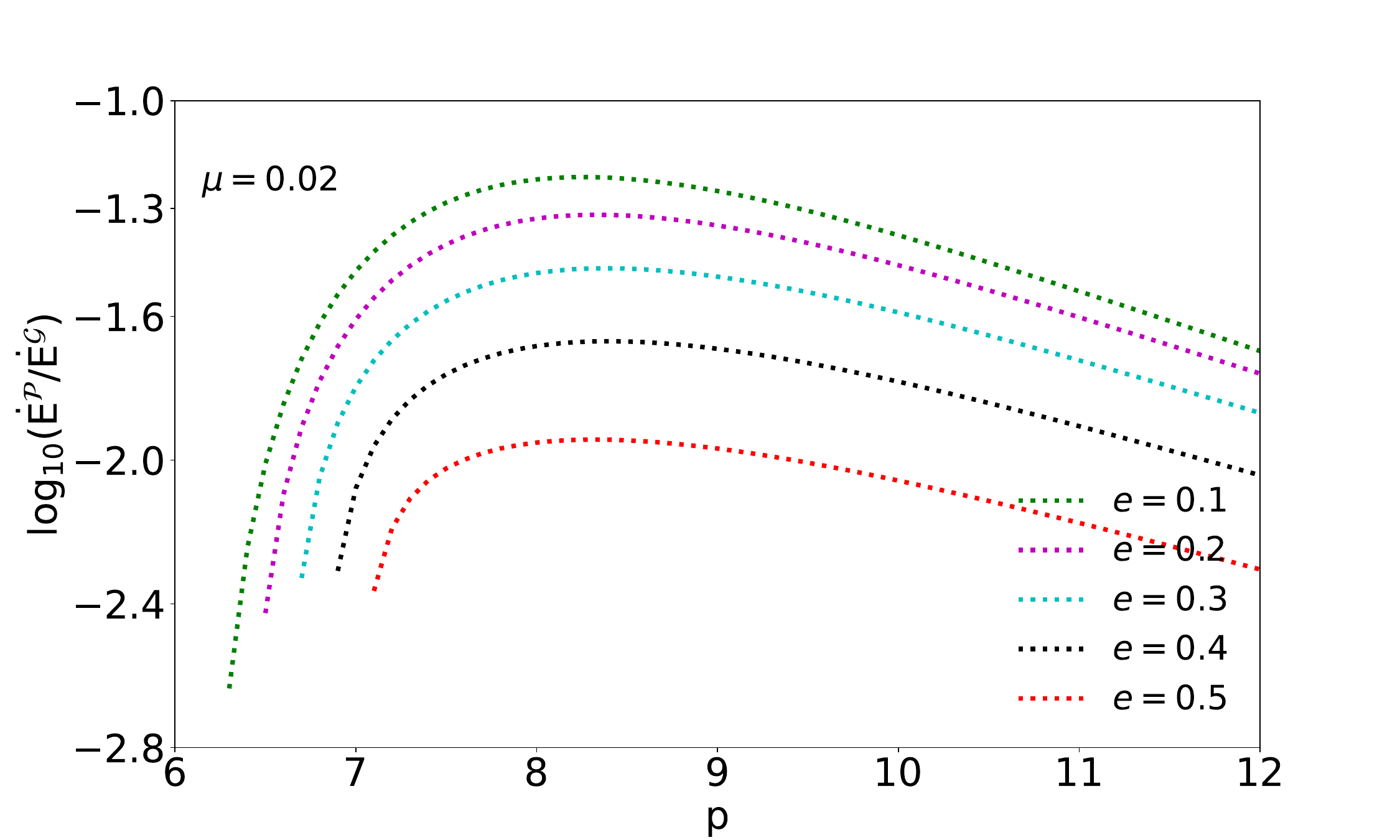}
\includegraphics[width=3.17in, height=2.3in]{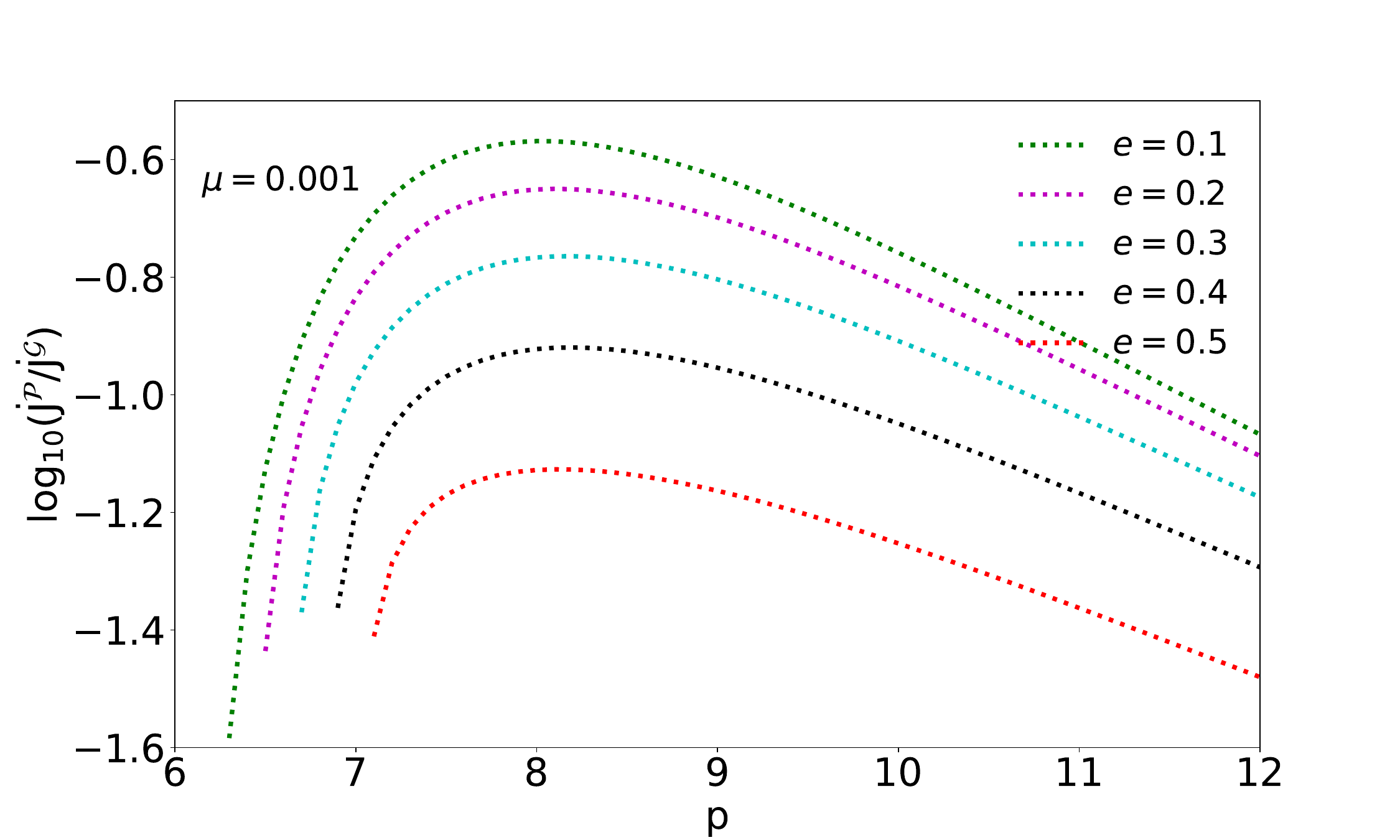}
\includegraphics[width=3.17in, height=2.3in]{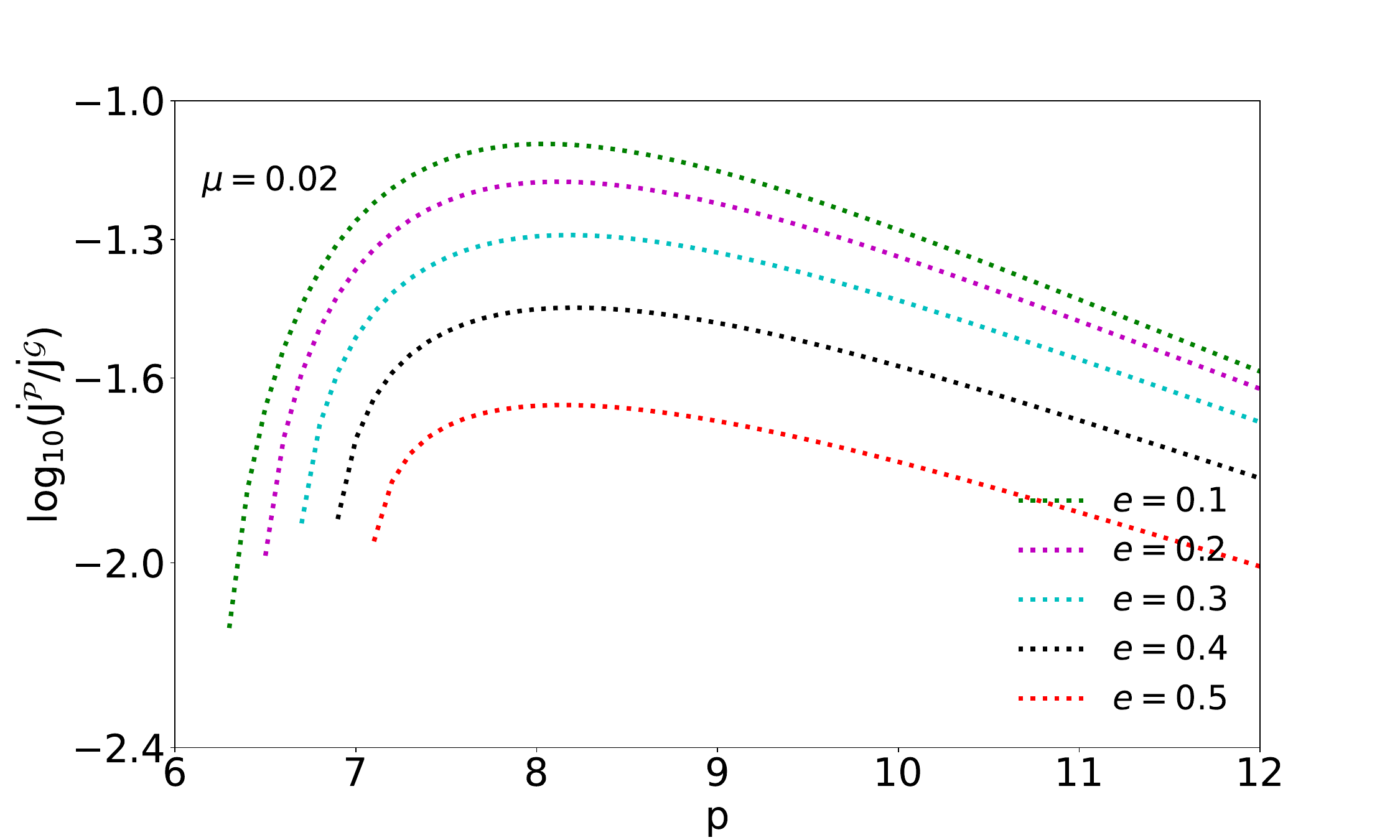}
\caption{Ratios of the Proca to gravitational fluxes as functions of the orbital semi-latus rectum $p$ for different initial eccentricities $e$. The left panels correspond to a Proca mass $\mu = 0.001$, and the right panels to $\mu = 0.02$. The energy fluxes are expressed in units of $m_p^2/M^2$, while the angular-momentum fluxes are expressed in units of $m_p^2/M$, where the primary mass is $10^6 M_\odot$ and the secondary mass is $10 M_\odot$.} \label{Fig:Fluxes}
\end{figure*}
\begin{figure*}[htb!]
\centering
\includegraphics[width=5.57in, height=3.2in]{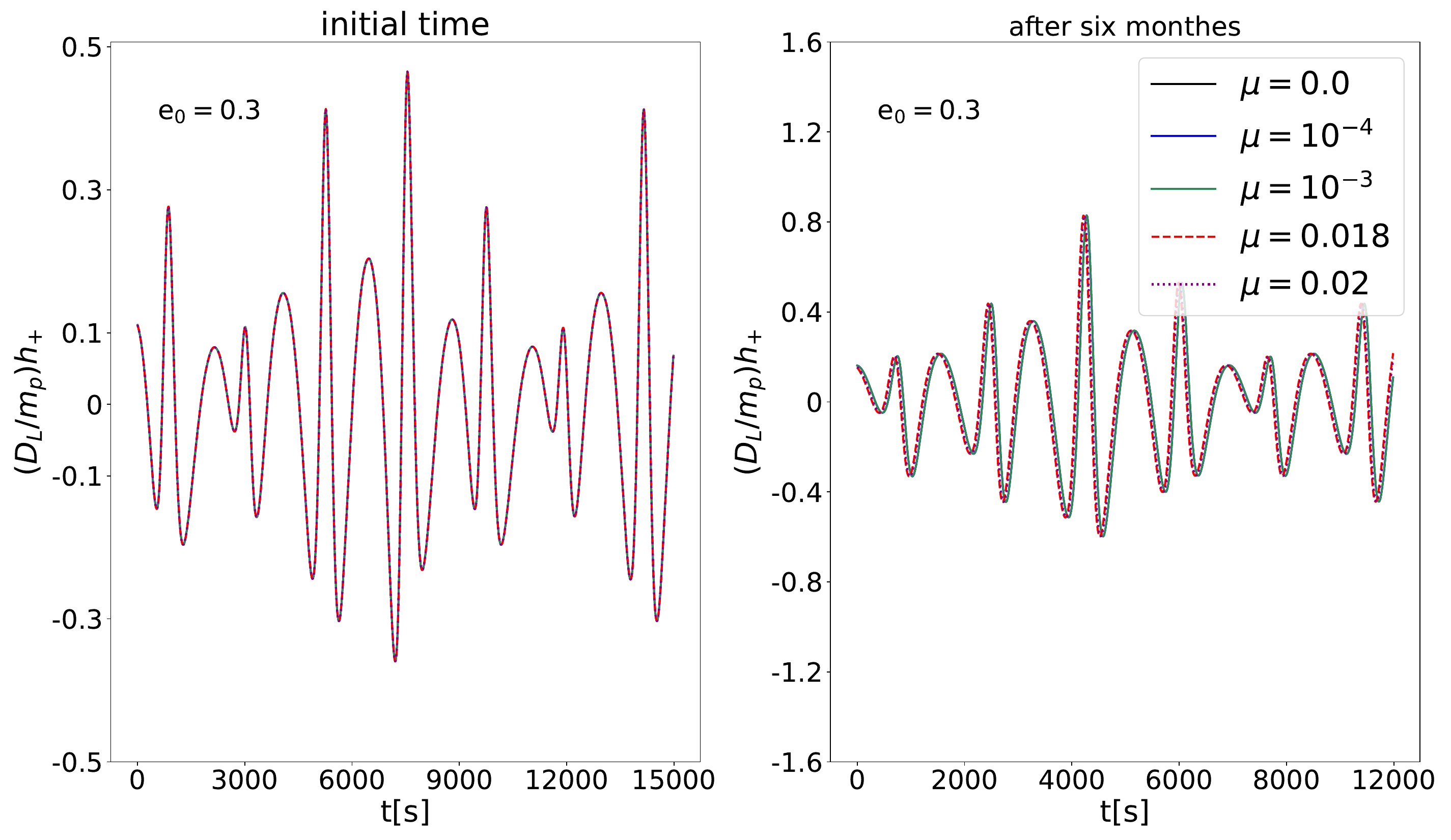}
\includegraphics[width=5.57in, height=3.2in]{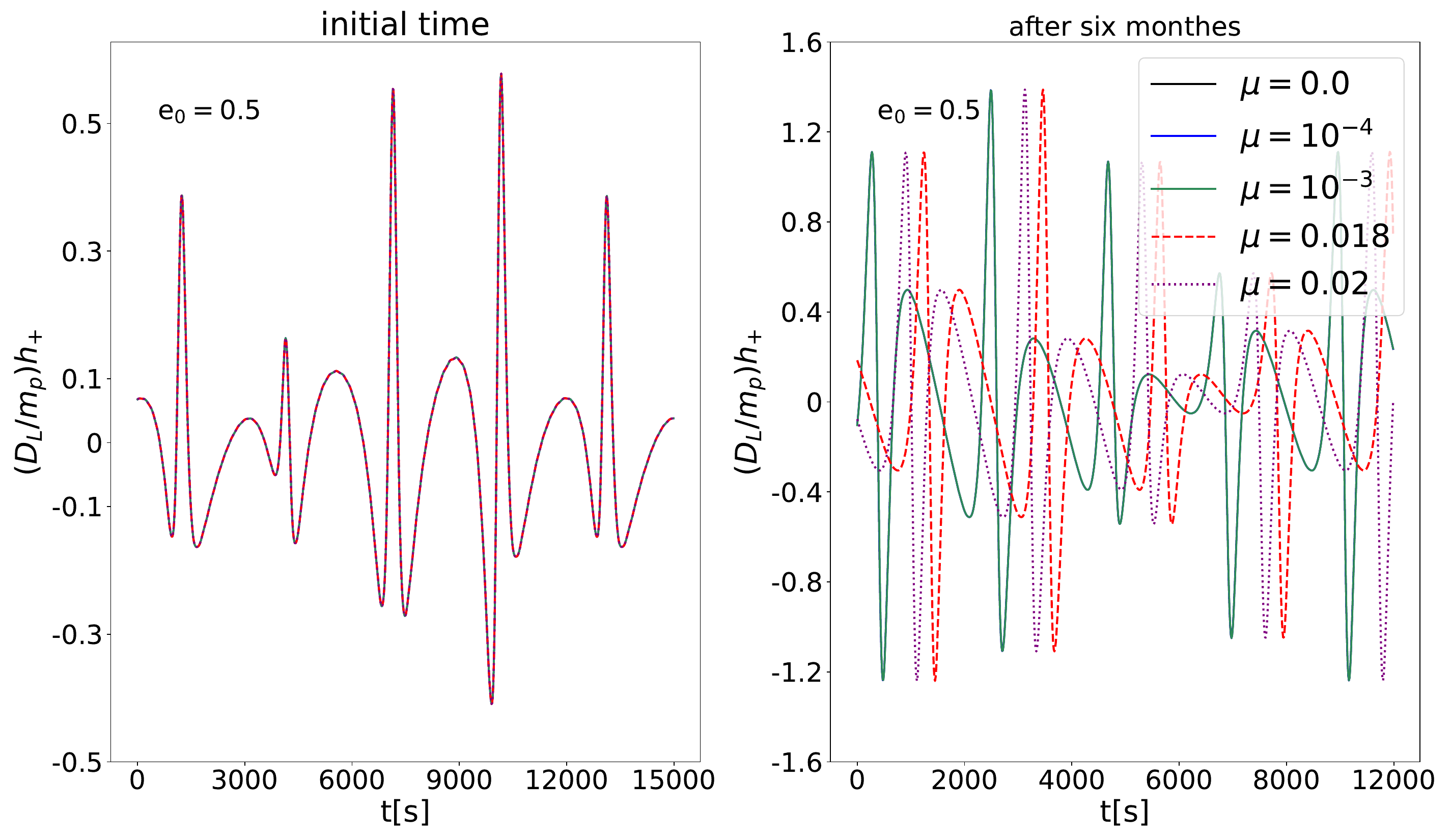}
\caption{Comparison of five waveforms from a total mass $10^6M_\odot+10M_\odot$ EMRIs system with the effect of the vector mass $\mu\in\{0.0,10^{-4},10^{-3},0.018,0.02\}$ are plotted for the initial eccentricity $e_0\in\{0.3,0.5\}$ and semi-latus rectum $p_0=12$.
} \label{Fig:wave1}
\end{figure*}

\begin{figure*}[htb!]
\centering
\includegraphics[width=3.17in, height=2.3in]{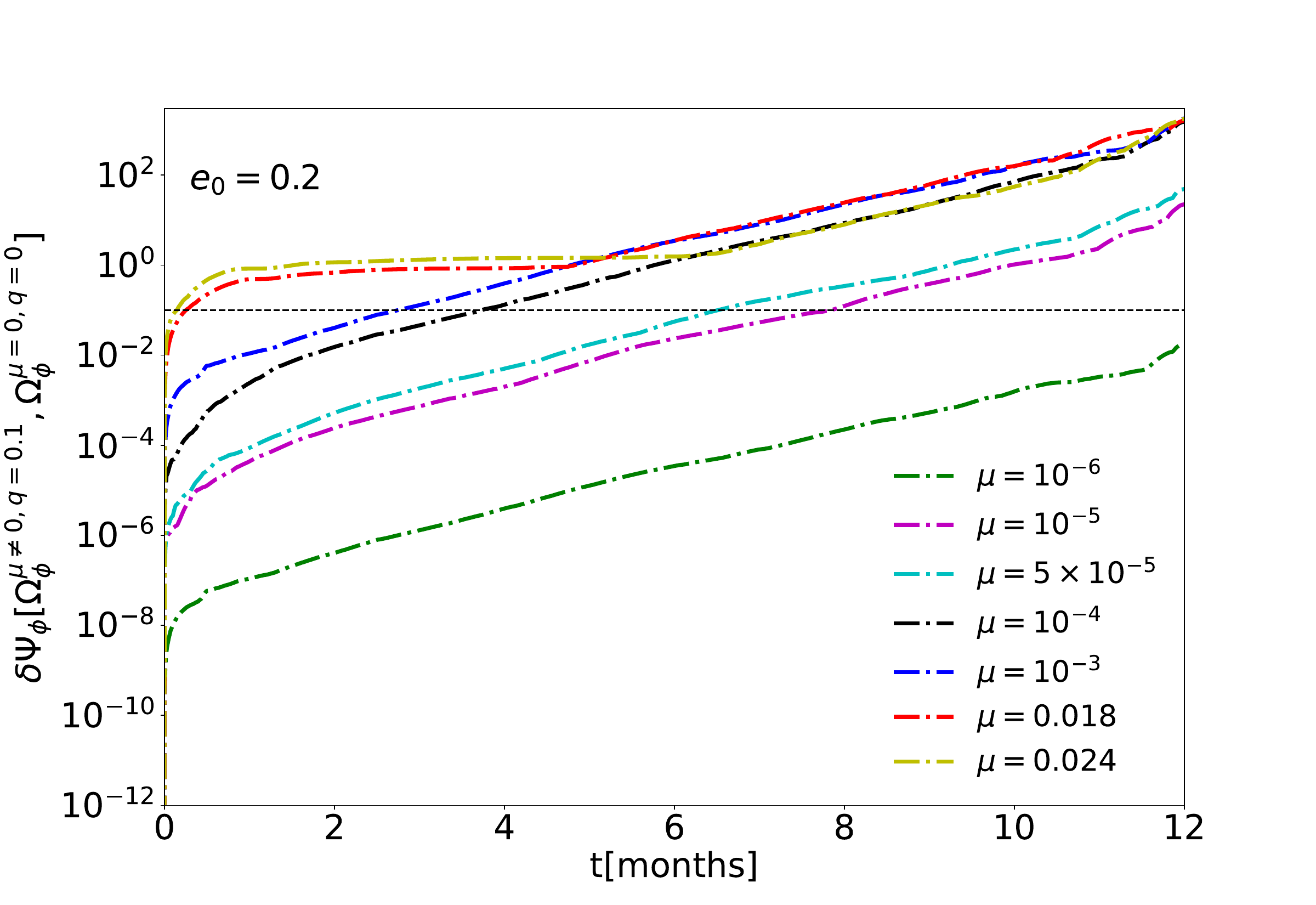}
\includegraphics[width=3.17in, height=2.3in]{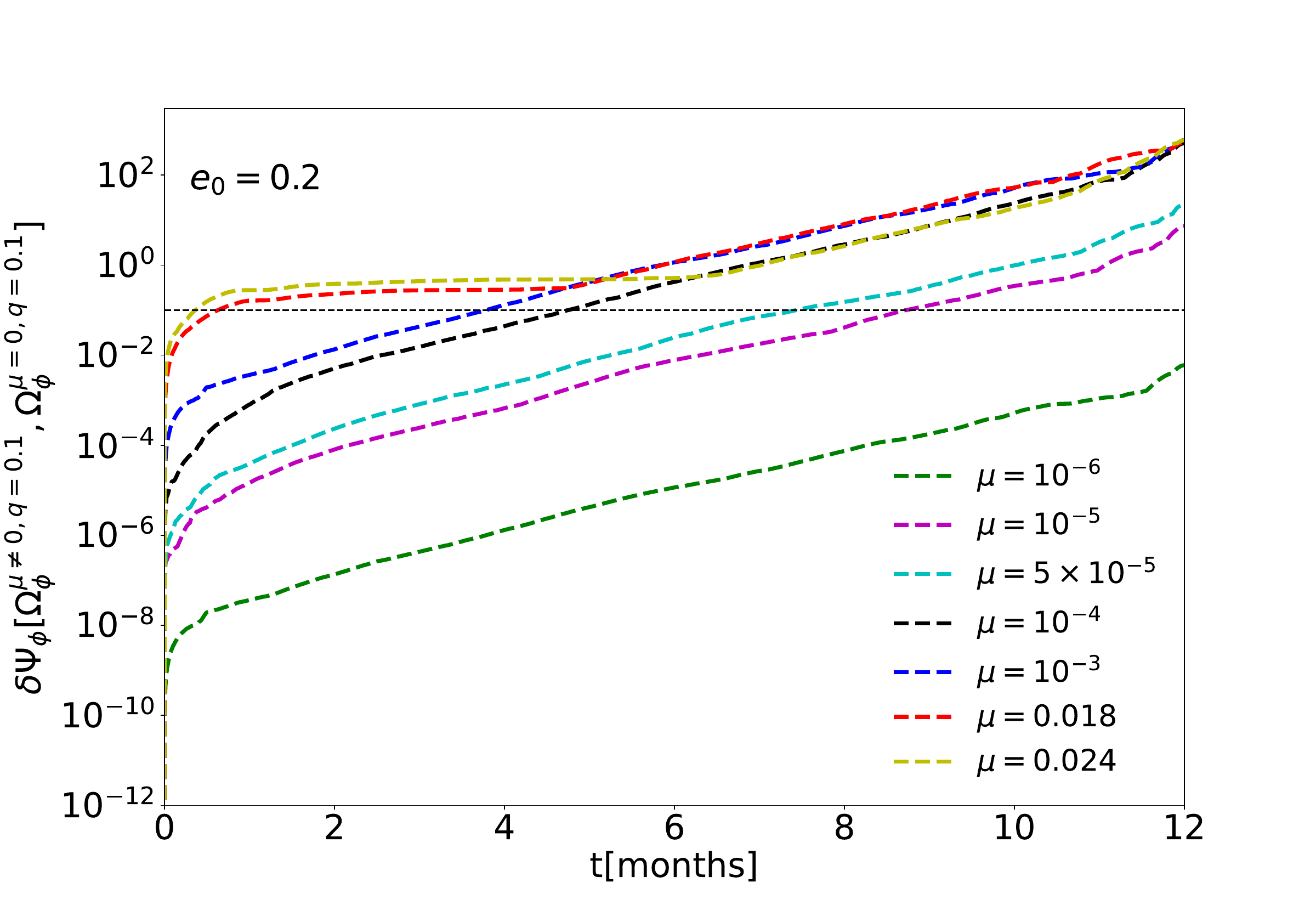}
\includegraphics[width=3.17in, height=2.3in]{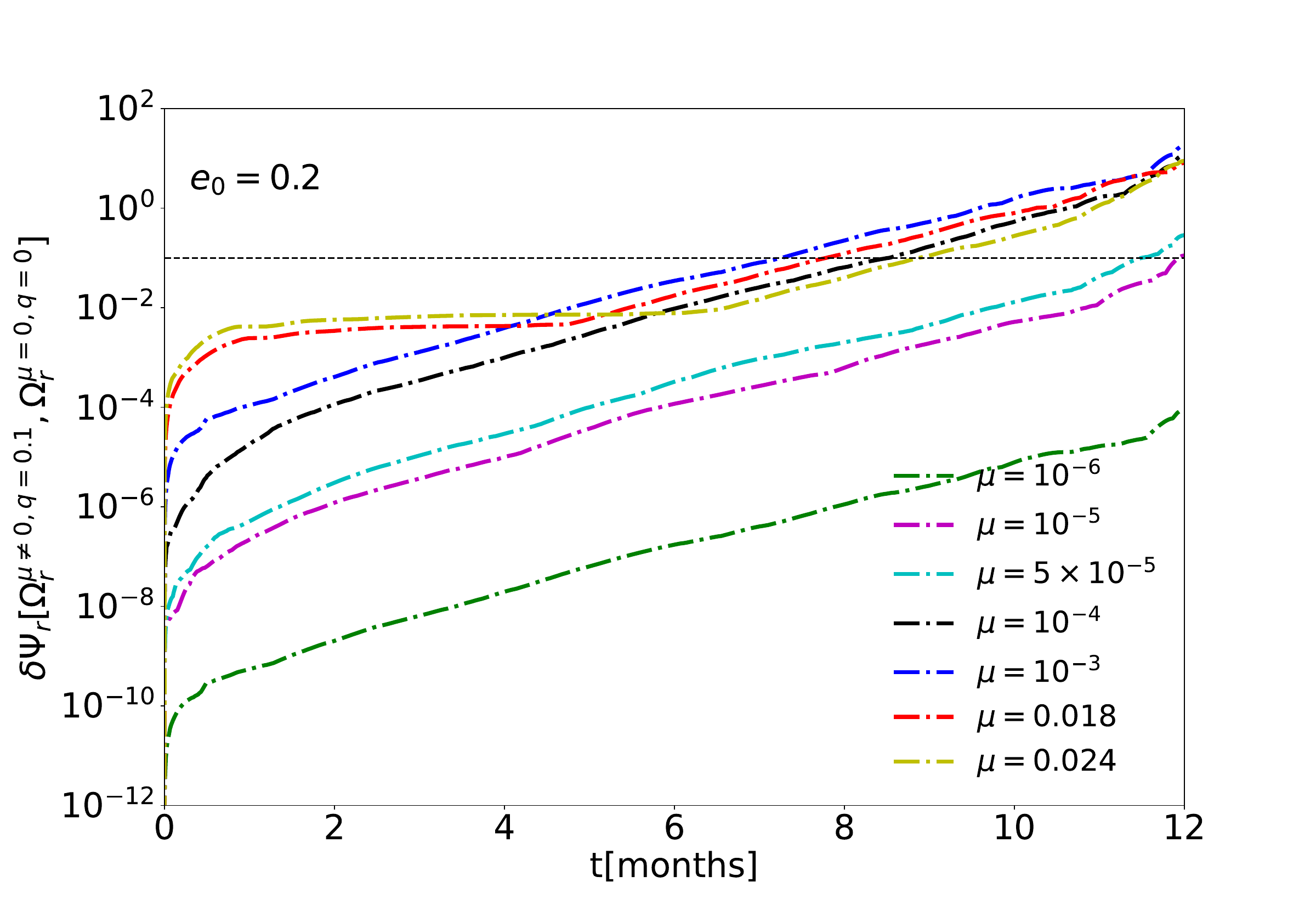}
\includegraphics[width=3.17in, height=2.3in]{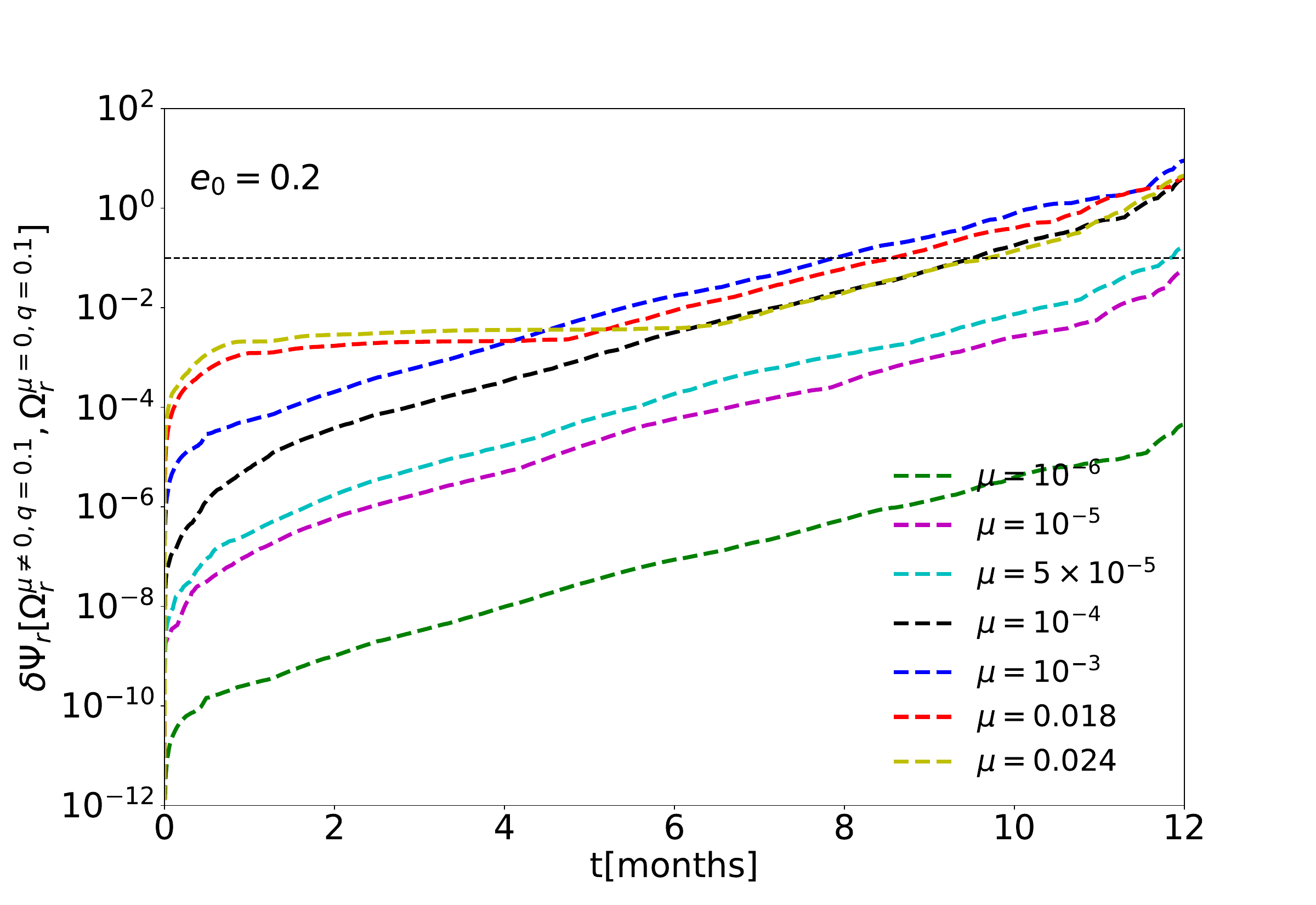}
\caption{Azimuthal (top panels) and radial (bottom) dephasing as a function of observation time for initial orbital eccentricity $e_0 =0.2$ and orbital semi-latus rectum $p_0=12.0$ is plotted, the other parameters are setting as following: $\mu \in\{10^{-6}, 10^{-5}, 5\times10^{-5}, 10^{-4}, 10^{-3},0.018,0.024\}$, vector charge $q=0.1$. The dephasing plots consider two cases:
contribution of Proca flux and massless vector flux on the evolution of radial or azimuthal frequency (left panels); contribution of Proca flux and standand GR flux on the evolution of radial or azimuthal frequency  (right panels).
The horizontal black dashed line in four figures denotes to the threshold distinguished by LISA, where the SNR of EMRIs signal is 30 \cite{Bonga:2019ycj}, other parameters keep same with Fig.~\ref{Fig:Fluxes}.} \label{Fig:dephaing1}
\end{figure*}

\begin{figure*}[htb!]
\centering
\includegraphics[width=3.17in, height=2.3in]{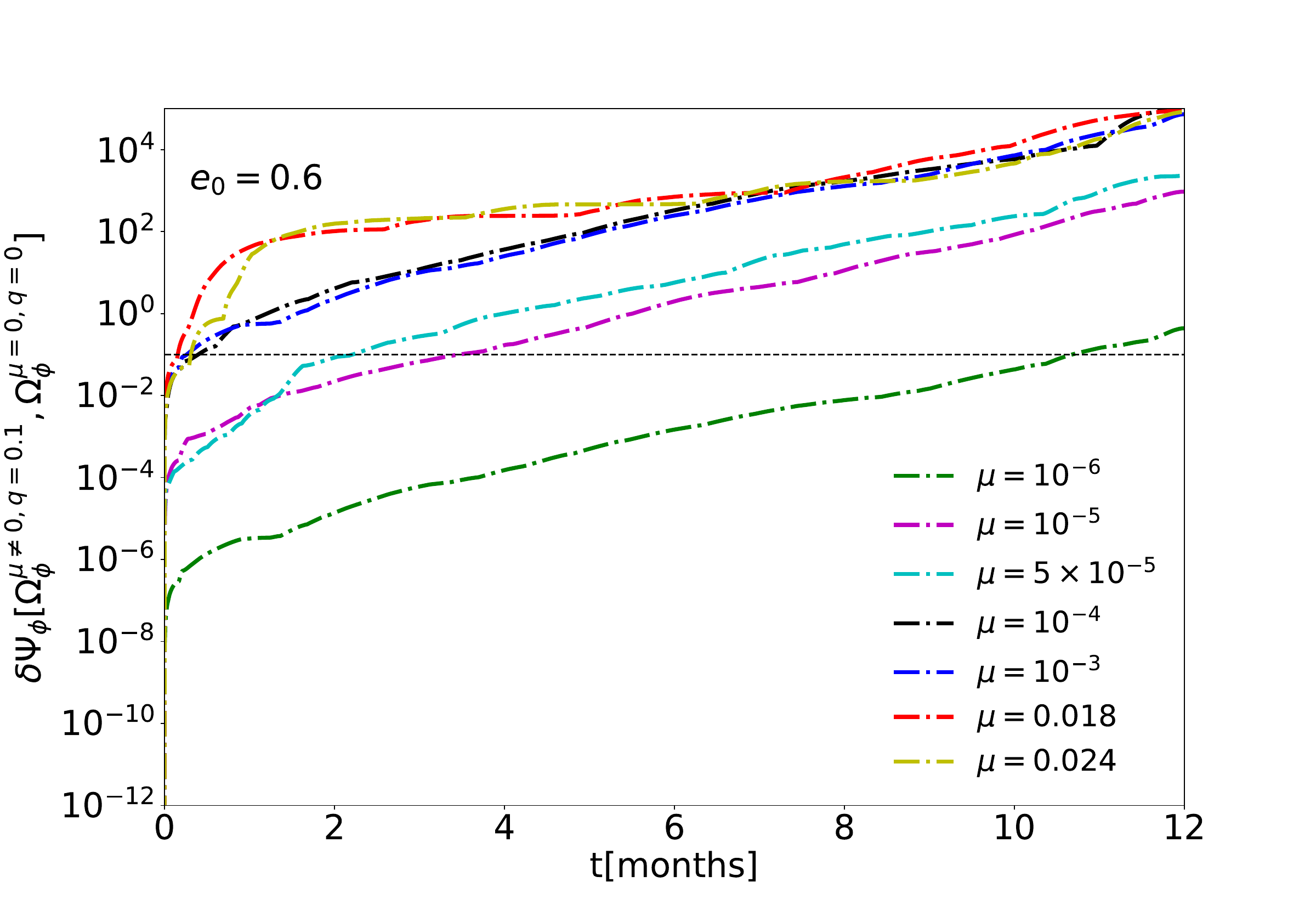}
\includegraphics[width=3.17in, height=2.3in]{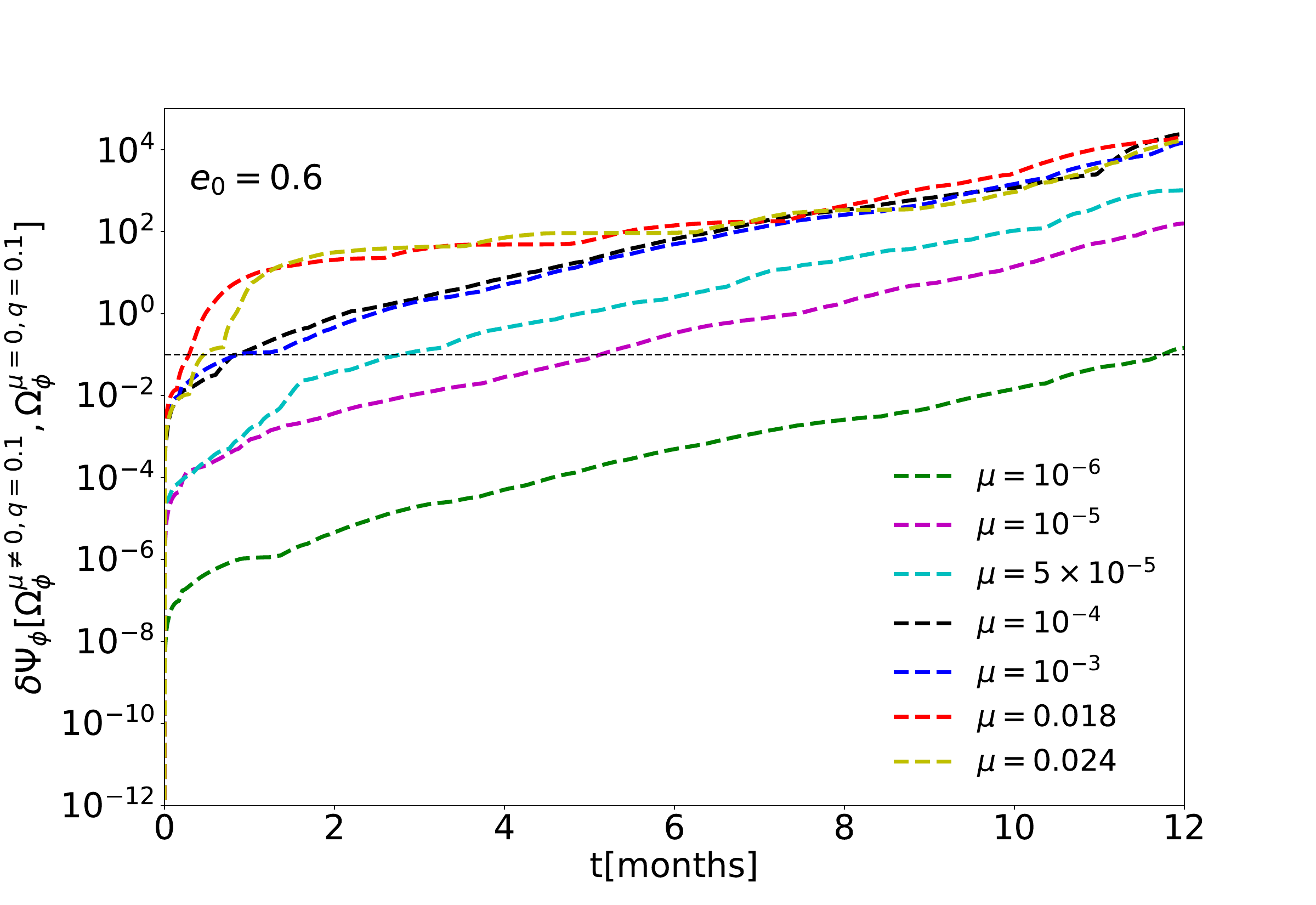}
\includegraphics[width=3.17in, height=2.3in]{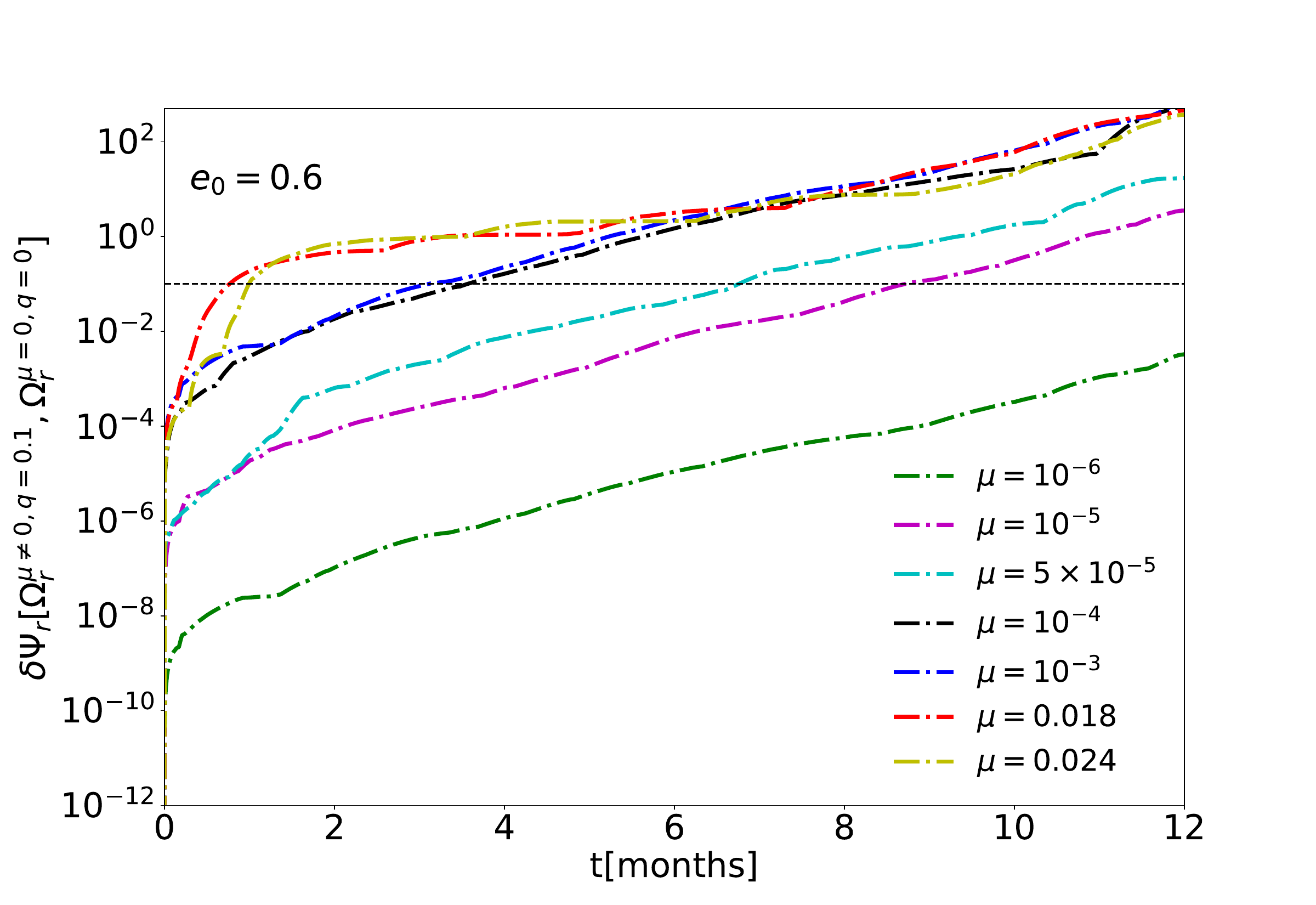}
\includegraphics[width=3.17in, height=2.3in]{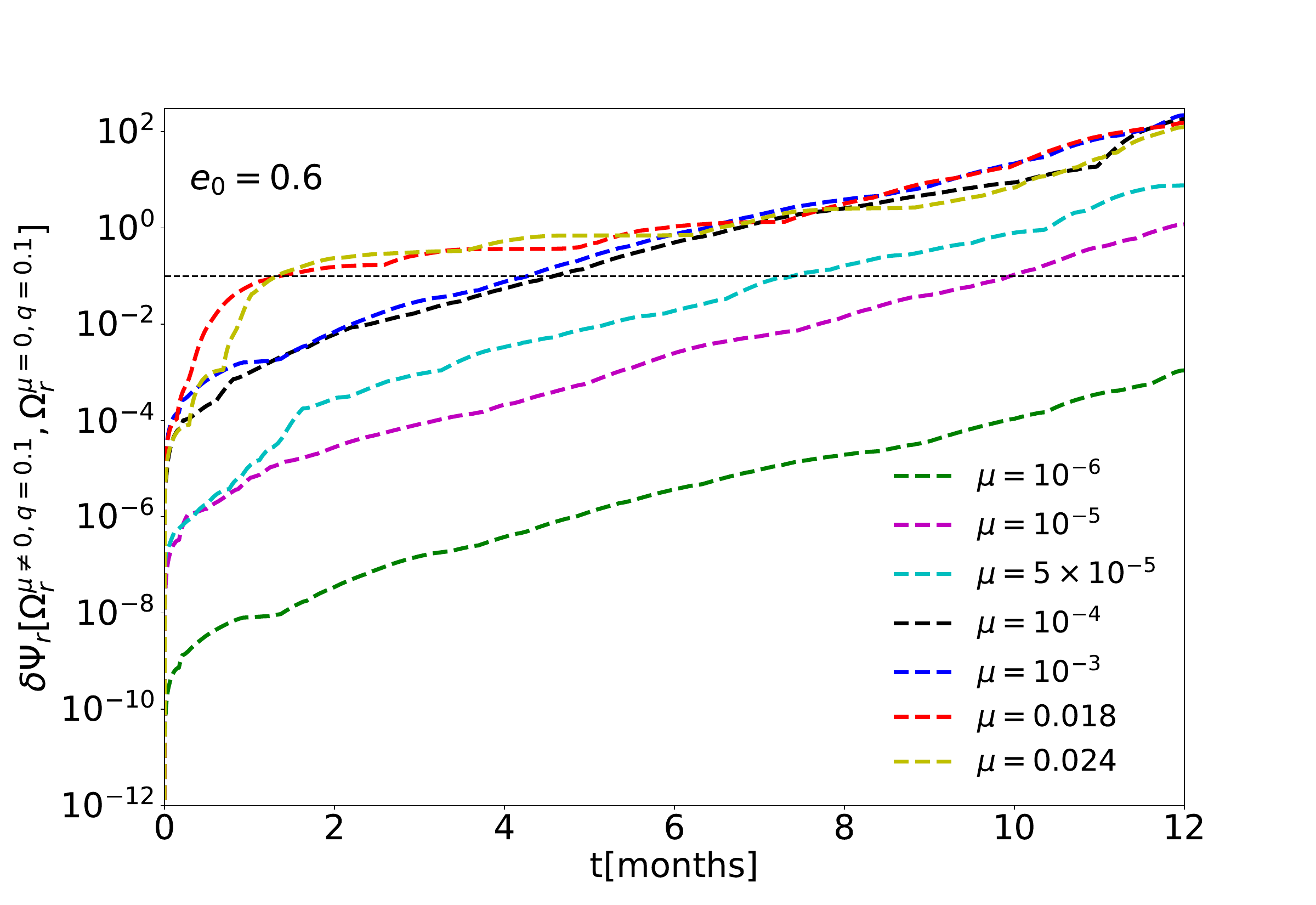}
\caption{Azimuthal (top panels) and radial (bottom panels) dephasing as a function of observation time for initial orbital eccentricities $e_0 =0.6$ are plotted, the configuration of other parameters keeps same with Fig.~\ref{Fig:dephaing1}.} \label{Fig:dephaing2}
\end{figure*}

\begin{figure*}[htb!]
\centering
\includegraphics[width=3.17in, height=2.3in]{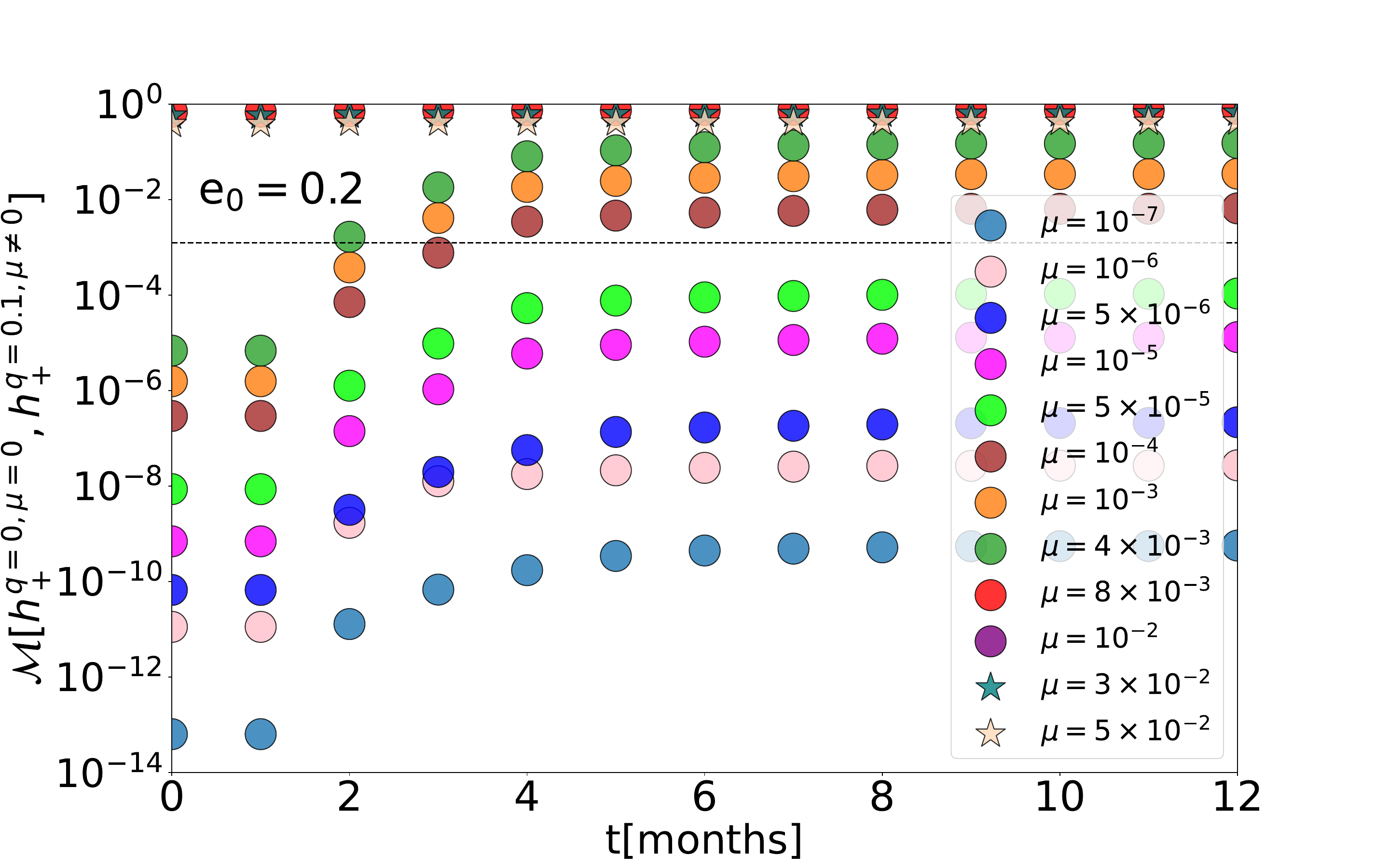}
\includegraphics[width=3.17in, height=2.3in]{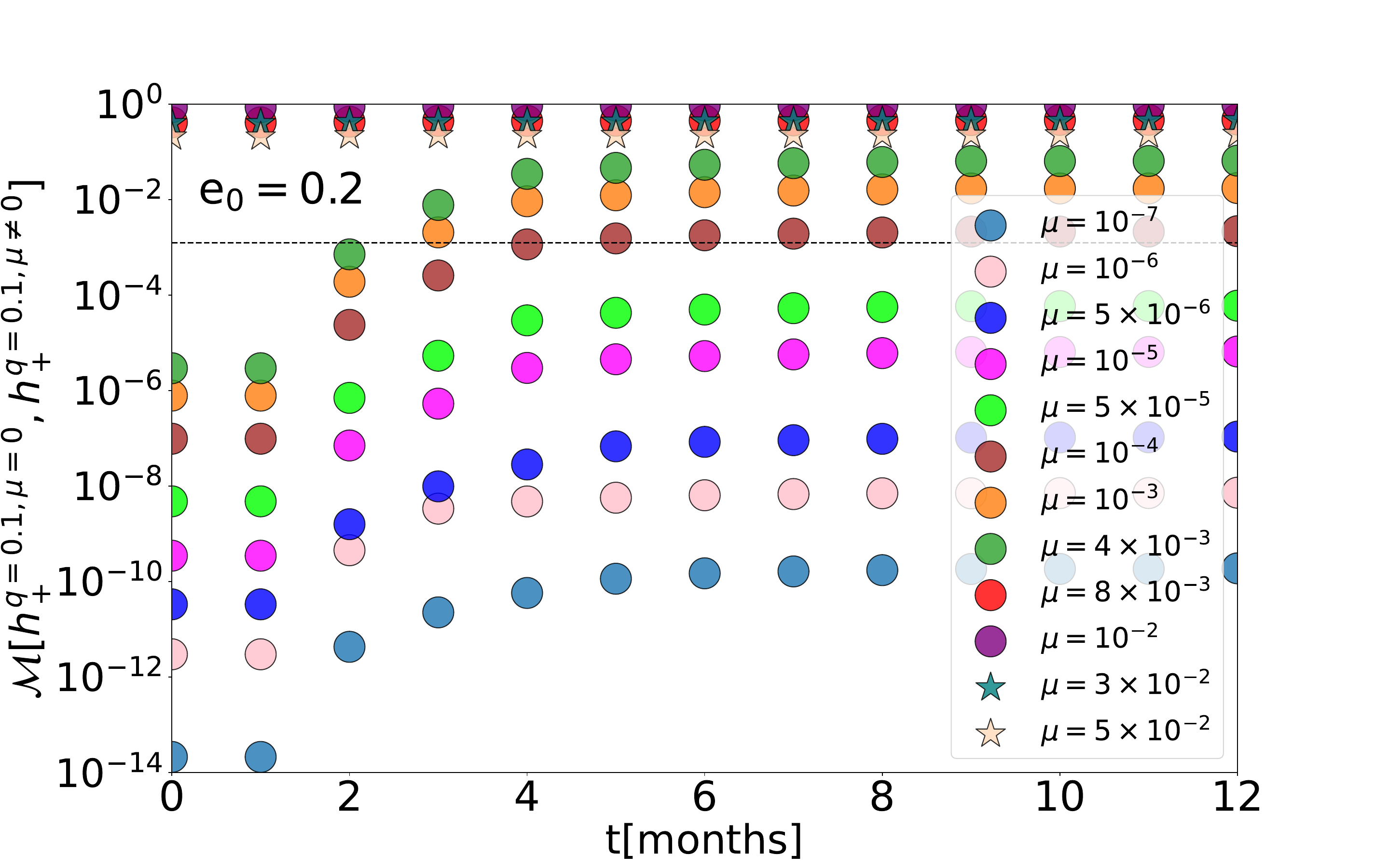}
\includegraphics[width=3.17in, height=2.3in]{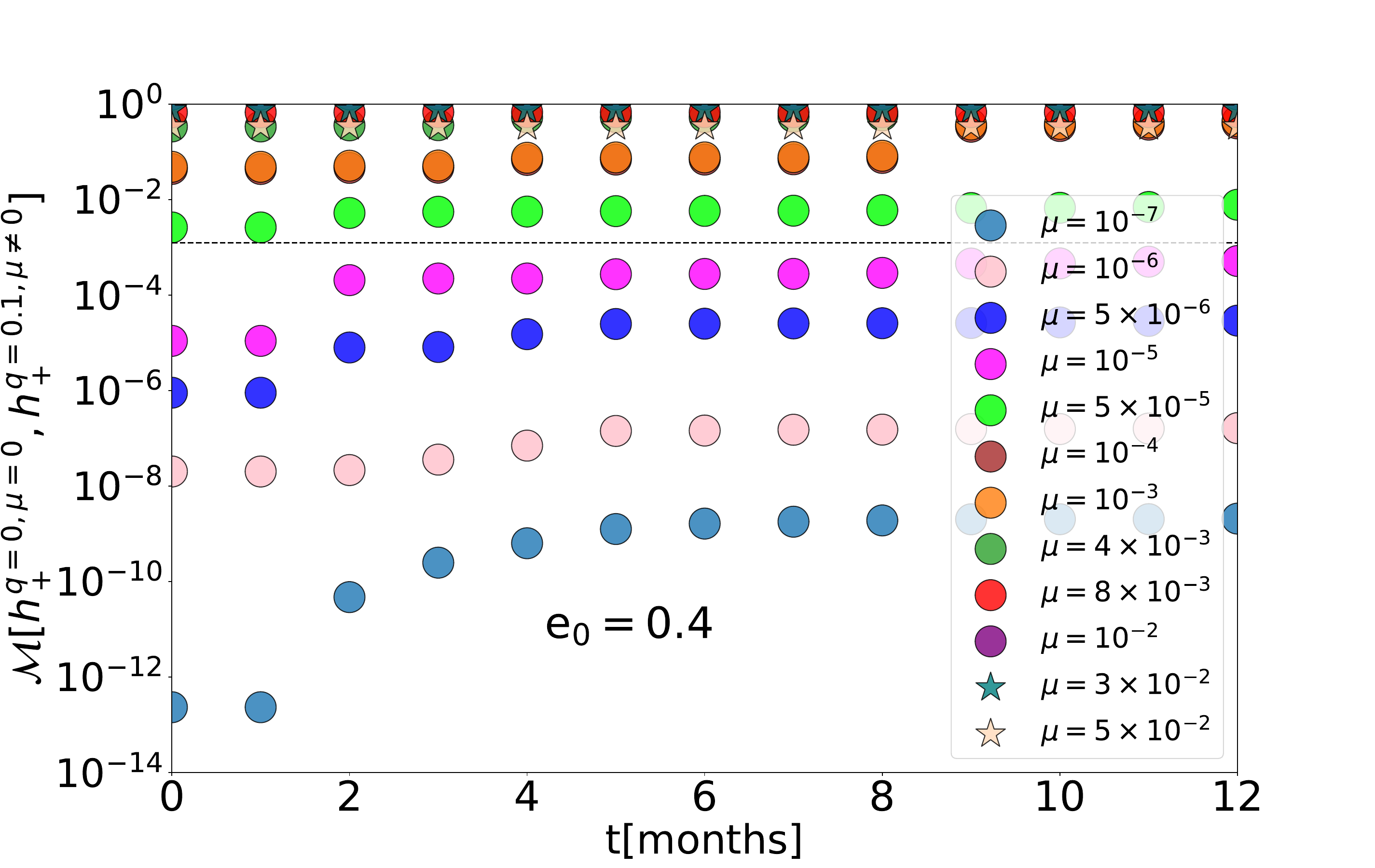}
\includegraphics[width=3.17in, height=2.3in]{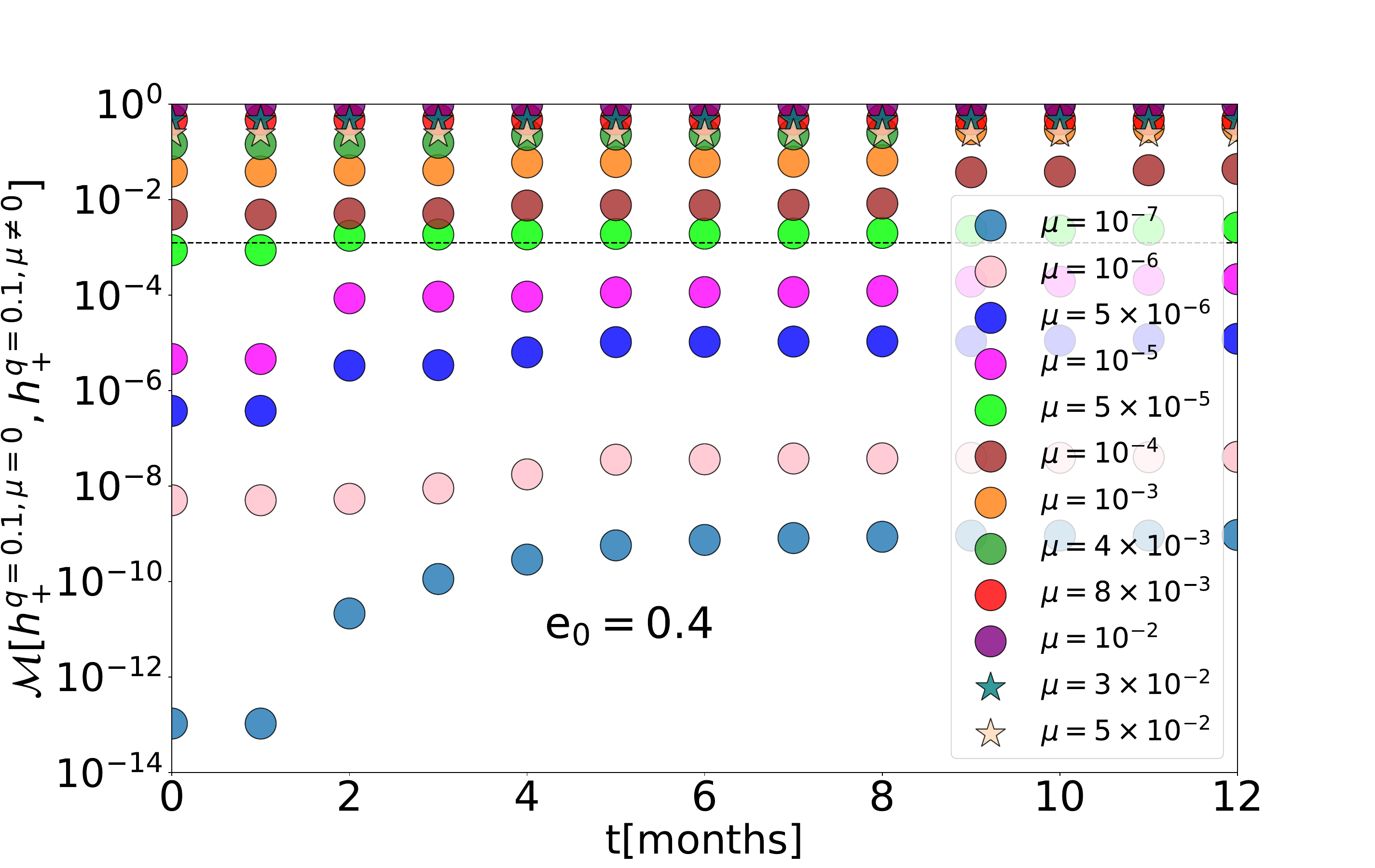}
\caption{Mismatches as the functions of observation time for the initial orbital eccentricities $e_0=(0.2,0.4)$ are plotted, which consider four cases: the difference of EMRIs waveforms between the standard GR case (top left panel) or massive (top right panel)  and massless vector field for the eccentricity $e_0=0.2$; the bottom panels are cases of initial eccentricity $e_0=0.4$. Note that EMRIs waveforms contain the correction of Proca field with masses $\mu\in\{10^{-7},10^{-6},5\times 10^{-6},10^{-5},5\times 10^{-5},10^{-4},10^{-3},4\times 10^{-3},8\times {-3},10^{-2},3\times10^{-2},5\times10^{-2}\}$. The other parameters keep same with Fig.~\ref{Fig:dephaing1}. } \label{Fig:mismatch}
\end{figure*}

\begin{figure*}[htb!]
\centering
\includegraphics[width=3.17in, height=2.3in]{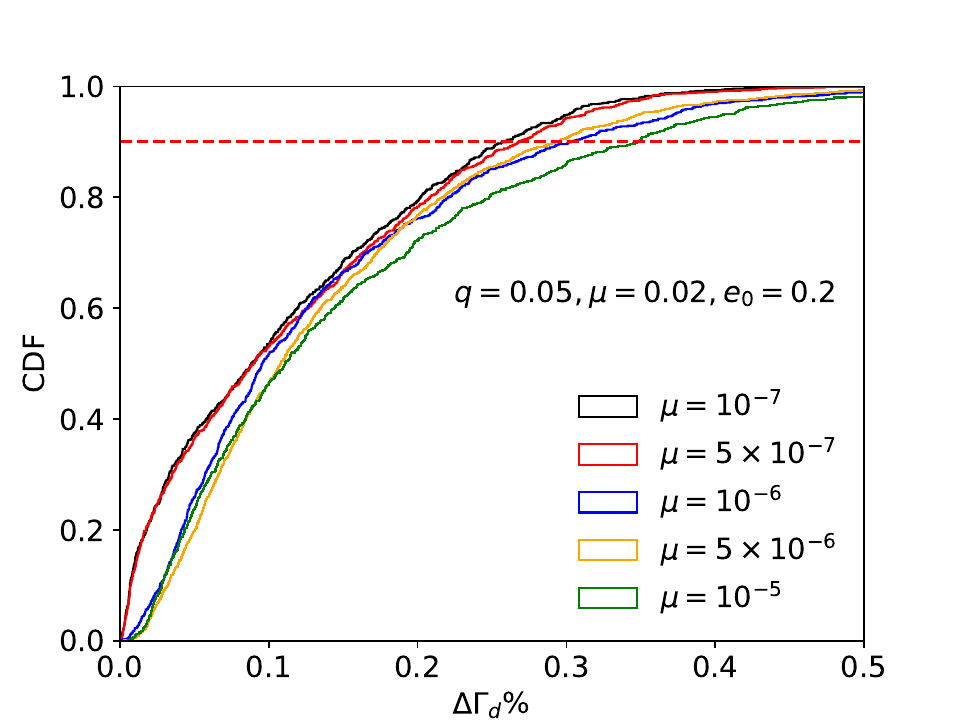}
\includegraphics[width=3.17in, height=2.3in]{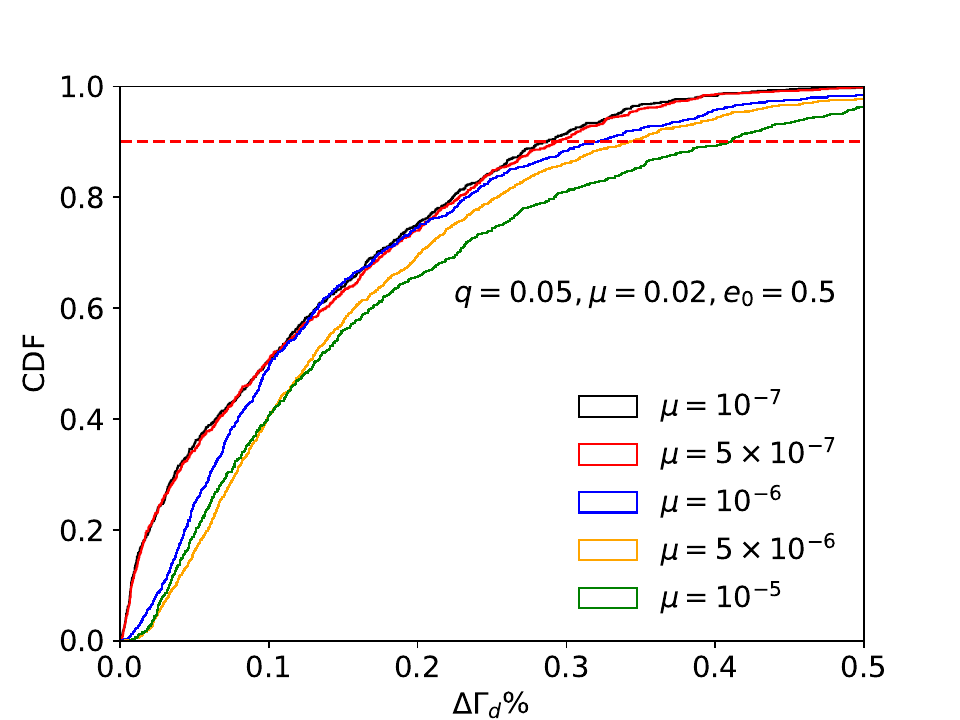}
\caption{
Cumulative distribution of the maximum relative error between the FIM and its perturbed counterpart, where each element of the deviation matrix follows a uniform distribution $u \in [-10^{-3},10^{-3}]$. Different colored curves correspond to distinct numerical derivative spacings used in the computation of the FIM. The left panel shows the case with orbital eccentricity $e_0 = 0.2$, with other parameters setting as in Fig.~\ref{fig7:cornerplot:ecc02}. The right panel corresponds to the case with $e_0 = 0.5$ keeping other parameters same with Fig.~\ref{fig8:cornerplot:ecc05}.
} \label{Fig:CDF}
\end{figure*}

Finally, we outline the procedure for estimating EMRI parameters and constraining the Proca mass. We employ the Fisher Information Matrix (FIM) formalism to place preliminary bounds on the Proca field mass. In gravitational-wave data analysis, the FIM is widely used to estimate the statistical uncertainties of source parameters, under the assumption of stationary, Gaussian noise and a sufficiently high signal-to-noise ratio (SNR) with flat priors \cite{Vallisneri:2007ev}. Under these conditions, the posterior distribution of the source parameters can be approximated by a multivariate normal distribution centered on the true values, with the diagonal elements of the covariance matrix representing the corresponding measurement errors. The FIM is defined in terms of the partial derivatives of the waveform with respect to each parameter
\begin{equation}
\boldsymbol{\Gamma}_{ij} = \left(\frac{\partial h(f)}{\partial\boldsymbol{\theta}_i} \Big|\frac{\partial h(f)}{\partial\boldsymbol{\theta}_j}\right)\;,
\end{equation}
where source parameters are $\boldsymbol{\theta}_i=\{M,m_p,p_0,e_0,\mu,q,\Phi_{r,0},
\Phi_{\phi,0},\theta_S,\phi_S,\theta_K,\phi_K,D_L\}$ with $i=1,2,\dots,13$, and
its inverse is the covariance matrix
\begin{equation}
\boldsymbol{\Sigma}_{\boldsymbol{\theta}_i \boldsymbol{\theta}_j} =
<\delta \boldsymbol{\theta}_i  \delta \boldsymbol{\theta}_j>
= (\boldsymbol{\Gamma}^{-1})_{ij}  \; . \label{eq:cov:matrix}
\end{equation}
Therefore, the uncertainty measured by detector can be approximately obtained by diagonal of covariance matrix \cite{Vallisneri:2007ev}
\begin{eqnarray}
\sigma_{\boldsymbol{\theta}_i} =
\sqrt{\boldsymbol{\Sigma}_{\boldsymbol{\theta}_i \boldsymbol{\theta}_i}} \label{sigma:fim}\;.
\end{eqnarray}
This method, in fact, can present a more rigorous bound on the mass of Proca field comparing to the
results obtained from dephasing and mismatch. Moreover, the off-diagonal elements of variance matrix can
quantify the correlation among the source parameters.

\section{Result}\label{result}
In this section, we begin with the comparison of Proca and gravitational fluxes. Figure~\ref{Fig:Fluxes} shows the ratios of the energy and angular-momentum fluxes as functions of the orbital semi-latus rectum for several initial eccentricities. Two representative EMRI systems with Proca masses $\mu=0.001$ (left panels) and $\mu=0.02$ (right panels) are considered, with eccentricities $e \in \{0.1,0.2,0.3,0.4,0.5\}$.
As shown in Fig.~\ref{Fig:Fluxes}, the flux ratio exhibits a non-monotonic dependence on the orbital separation. For fixed eccentricity, the ratio increases steadily as the binary inspirals, reaches a maximum around $p \sim 8$, and then decreases sharply as the orbit approaches the last stable orbit.

Next, we compare EMRI waveforms in the presence of different Proca masses.
Figure~\ref{Fig:wave1} shows waveforms for binaries with $\mu \in \{0,10^{-4},10^{-3},0.018,0.02\}$ and vector charge $q=0.1$, with initial eccentricities $e_0=0.3$ (top panels) and $e_0=0.5$ (bottom panels). We consider an EMRI system with total mass $10^6 M_\odot + 10 M_\odot$, where the secondary undergoes one year of inspiral.

As illustrated in the left panels of Fig.~\ref{Fig:wave1}, the five waveforms are initially indistinguishable, even at higher eccentricity ($e_0=0.5$). However, after six months of inspiral (right panels), phase differences begin to accumulate. For $e_0=0.3$, the phases remain nearly identical across all Proca masses, while for $e_0=0.5$, significant deviations appear. In particular, the phase difference grows with increasing Proca mass. Interestingly, for the heaviest case ($\mu=0.02$), the deviation is smaller than expected, suggesting that the influence of the Proca field on the waveform may be suppressed at higher vector masses.

\subsection{Dephasing and mismatch}
In this subsection, we present the results including dephasing and mismatch to assess the effect of Proca mass on EMRIs waveform.

In Fig. \ref{Fig:dephaing1}, we show the dephasing of EMRIs orbital phases between the Proca radiation and the massless vector or standard GR case for the initial eccentricities $e_0=0.2$.
The azimuthal dephasing is shown in the top panels, and the radial dephasing in the bottom panels, for Proca masses $\mu \in \{10^{-6}, 10^{-5}, 5\times10^{-5}, 10^{-4}, 10^{-3}, 0.018, 0.024\}$. The left panels display the phase differences between Proca radiation and GR, while the right panels show the difference between massive and massless vector-field cases. Here, the superscripts $\mu \in \{=0,\neq0\}$ and $q \in \{=0,\neq0\}$ indicate whether the orbital frequencies are evolved using Proca, massless-vector, or standard GR fluxes. For example, $\Omega_r^{\mu=0, q=0.1}$ denotes the radial frequency evolution with massless-vector fluxes for vector charge $q=0.1$.

Overall, the dephasing induced by the Proca field becomes more pronounced as the vector mass increases. However, beyond a certain threshold, the dephasing slightly decreases, for both azimuthal and radial components. A similar trend is observed for higher initial eccentricity ($e_0=0.6$) in Fig.~\ref{Fig:dephaing2}. Comparing the two cases, we find that dephasing is systematically larger for $e_0=0.6$ than for $e_0=0.2$. This indicates that higher orbital eccentricity enhances the detectability of Proca-field effects in EMRIs with LISA.

Finally, in Fig.~\ref{Fig:mismatch}, we present the mismatch as a function of observation time for initial orbital eccentricities $e_0 \in \{0.2, 0.4\}$, comparing the Proca radiation with standard general relativity (GR) and with massless vector radiation. The horizontal black dashed line marks the mismatch threshold $\mathcal{M} \sim 10^{-3}$, above which two EMRI signals can be distinguished by the LISA detector.
The top panels correspond to $e_0=0.2$, displaying mismatches between Proca and GR waveforms (top left) and between Proca and massless-vector waveforms (top right) for a fixed vector charge $q=0.1$ and varying Proca mass $\mu \in \{10^{-7}, 10^{-6}, 5 \times 10^{-6}, 10^{-5}, 5 \times 10^{-5}, 10^{-4}, 10^{-3}, 4 \times 10^{-3}, 8 \times 10^{-3}, 10^{-2}, 3 \times 10^{-2}, 5 \times 10^{-2}\}$. The bottom panels show the corresponding results for $e_0=0.4$.

Across all panels, the mismatch generally increases with growing vector mass $\mu$, indicating stronger deviations from the reference waveforms. However, once the Proca mass becomes sufficiently large, the mismatch begins to decrease. This damped trend of mismatch arises because higher Proca masses suppress the radiated flux, which is also reflected in Table~\ref{Tab:Fluxvalues}.
Moreover, the mismatches between the Proca flux modified and GR waveforms are consistently larger than those between Proca and massless vector waveforms. This is expected, as the former comparison encapsulates differences in both vector charge $q$ and mass $\mu$, whereas the latter includes the effect of $q$ alone, holding the massless vector case as reference. Additionally, increasing orbital eccentricity leads to slightly larger mismatches, since more eccentric EMRIs encode richer waveform structures and allow for more prominent differences in radiation signatures.
Finally, for a fixed vector charge $q = 0.1$, the minimum discernible Proca mass by LISA is approximately $\mu\sim 10^{-4} ~(\sim1.33\times10^{-20} \rm eV)$, while for higher eccentricity $e_0 = 0.4$, the EMRIs signal would carry more information about Proca field, allowing distinguishing of vector masses as low as $\mu\sim 5\times10^{-5} ~(\sim6.67\times10^{-21} \rm eV)$.
Therefore, the eccentricity of EMRIs orbits play a key role in the constraint on Proca mass, we should model the Proca fluxes for eccentric insprials  with great deliberation.

\begin{table*}[htbp!]
\centering
\begin{tabular}{cc|ccccccccccccc}
\hline
\hline
$\mu$ & $e$ & $\sigma_M/M$ & $\sigma_{m_p}/m_p$   &$\sigma_{p_0}$
&$\sigma_{e_0}$  & $\sigma_{\mu}/\mu$  & $\sigma_{q}$
& $\sigma_{\theta_S}/\theta_S$ & $\sigma_{\phi_S}/\phi_S$
 & $\sigma_{\theta_K}/\theta_K$   & $\sigma_{\phi_K}/\phi_K$
& $\sigma_{\Phi{_{\phi,0}}}/\Phi_{\phi,0}$
 & $\sigma_{\Phi{_{r,0}}}/\Phi_{r,0}$  & $\sigma_{D_L}$
\\
\hline
0.01  & 0.0  &$1.90\text{e-5}$  &$3.27\text{e-4}$  &$1.26\text{e-4}$ &$-$ & $774\%$
& $3.86\text{e-3}$  &$0.86$  & $6.36\text{e-1}$
&$7.57\text{e-1}$ &$4.57$
&$8.78\text{e-1}$  &$8.54\text{e-1}$  &$8.24$
\\
  &0.1   &$8.67\text{e-6}$  &$2.75\text{e-4}$  &$4.43\text{e-4}$ &$3.25\text{e-4}$ & $462\%$
&$3.56\text{e-3}$  &$0.76$  &$5.42\text{e-1}$
&$6.43\text{e-1}$ &$2.13$
&$8.34\text{e-1}$  &$7.57\text{e-1}$  &$7.68$
\\
 &0.5  &$3.23\text{e-6}$  &$1.30\text{e-5}$  &$2.45\text{e-5}$ &$1.70\text{e-4}$
& $121\%$   &$2.89\text{e-3}$  &$1.15\text{e-1}$
&$2.07\text{e-1}$  &$1.01\text{e-1}$ &$1.01\text{e-1}$
 &$1.48\text{e-1}$  &$3.18\text{e-1}$  &$3.22$
\\
\hline
\hline
0.02  & 0.0  &$2.50\text{e-5}$  &$5.26\text{e-4}$  &$1.57\text{e-4}$ &$-$ & $29.65\%$
& $6.43\text{e-3}$  &$0.76$  & $6.54\text{e-1}$
&$7.68\text{e-1}$ &$3.87$
&$8.15\text{e-1}$  &$7.78\text{e-1}$  &$8.45$
\\
&0.1   &$7.87\text{e-6}$  &$3.78\text{e-4}$  &$2.43\text{e-4}$ &$3.57\text{e-4}$ & $12.87\%$
&$3.56\text{e-3}$  &$0.68$  &$4.87\text{e-1}$
&$6.45\text{e-1}$ &$2.14$
&$8.26\text{e-1}$  &$7.45\text{e-1}$  &$6.78$
\\
&0.5  &$2.43\text{e-6}$  &$1.45\text{e-5}$  &$2.25\text{e-5}$ &$1.57\text{e-4}$
& $6.54\%$   &$2.78\text{e-3}$  &$1.55\text{e-1}$
&$1.89\text{e-1}$  &$1.51\text{e-1}$ &$1.31\text{e-1}$
 &$1.78\text{e-1}$  &$3.58\text{e-1}$  &$3.46$  \\
\hline
\hline
\end{tabular}
\caption{Measurement errors for EMRIs with masses $(M=10^{6}M_\odot,m_p= 10M_\odot)$,
vector charge and vector $(q=0.05,\mu\in\{0.0,0.01,0.02\}$, orbital semi-latus rectum $(p_0=12.0)$ and
eccentricity $e_0\in\{0,0.1,0.5\}$, initial orbital phases $(\Phi_{r,0}=\Phi_{\phi,0}=1.0)$ and
the directional angles related to source $(\theta_{S,K}=\phi_{S,K}=1.0)$ are listed, where the secondary object lasts the spiraling of two years and the luminosity distance $(D_L)$ is adjusted to set SNR of EMRIs signal as $150$. The transverse line below $\sigma_{e_0}$ means the circular EMRIs.
}\label{tab:fim:error}
\end{table*}

\subsection{Bound on mass of Proca field}
In this subsection, we assess the potential capacity of LISA on detecting Proca mass with the future observation of EMRI. Table~\ref{tab:fim:error} summarizes the parameter estimation of source's all parameters via FIM approach.
Note that to compare the constraint on vector mass with the circular and eccentric EMRIs, we adopt same vector charge configuration $(q=0.05)$ as that used in Ref.~\cite{Zi:2024lmt}.
The first and fourth rows report the measurement errors for the circular EMRIs, the other parameters configurations of Table II are setting as Ref.~\cite{Zi:2024lmt}.
Our results of parameter estimation show that EMRIs with higher orbital eccentricity yield tighter constraints on source parameters. In particular, the relative uncertainty on the Proca mass is substantially reduced from $774\%$ to $121\%$ for a Proca mass of $\mu = 0.01$. For signals incorporating a larger Proca mass, $\mu = 0.02$, LISA can measure the relative error to within $6.54\% (\sim 8.72\times10^{-18}\rm eV)$, demonstrating significantly improved  in precision.
This may be because the EMRI signal for a more massive vector field $(\mu=0.02)$ exhibits a larger deviation from the massless vector field case, the point that the vector flux is more different from the massless vector field case  is also be analyzed. Moreover, secondary object on a more eccentric orbit can inspiral deeper into the strong-field region, thereby carrying richer information about the source parameters.
Comparing to the results from circular EMRIs~\cite{Zi:2024lmt}, it is found that the measurement precision of intrinsic parameters (masses of binary, orbital parameters) would be improved slightly, and the extrinsic parameters (sky location and luminosity distance) do not display too much changing.
These results highlight the crucial role of orbital eccentricity in constraining deviations of EMRI signals from predictions of GR.
This enhancement may be due to that, for inspirals of comparable duration, the evolution of the semi-latus rectum in eccentric orbits exhibits greater deviations than in circular cases.

\begin{figure*}[htb!]

\centering
\includegraphics[width=0.87\paperwidth]{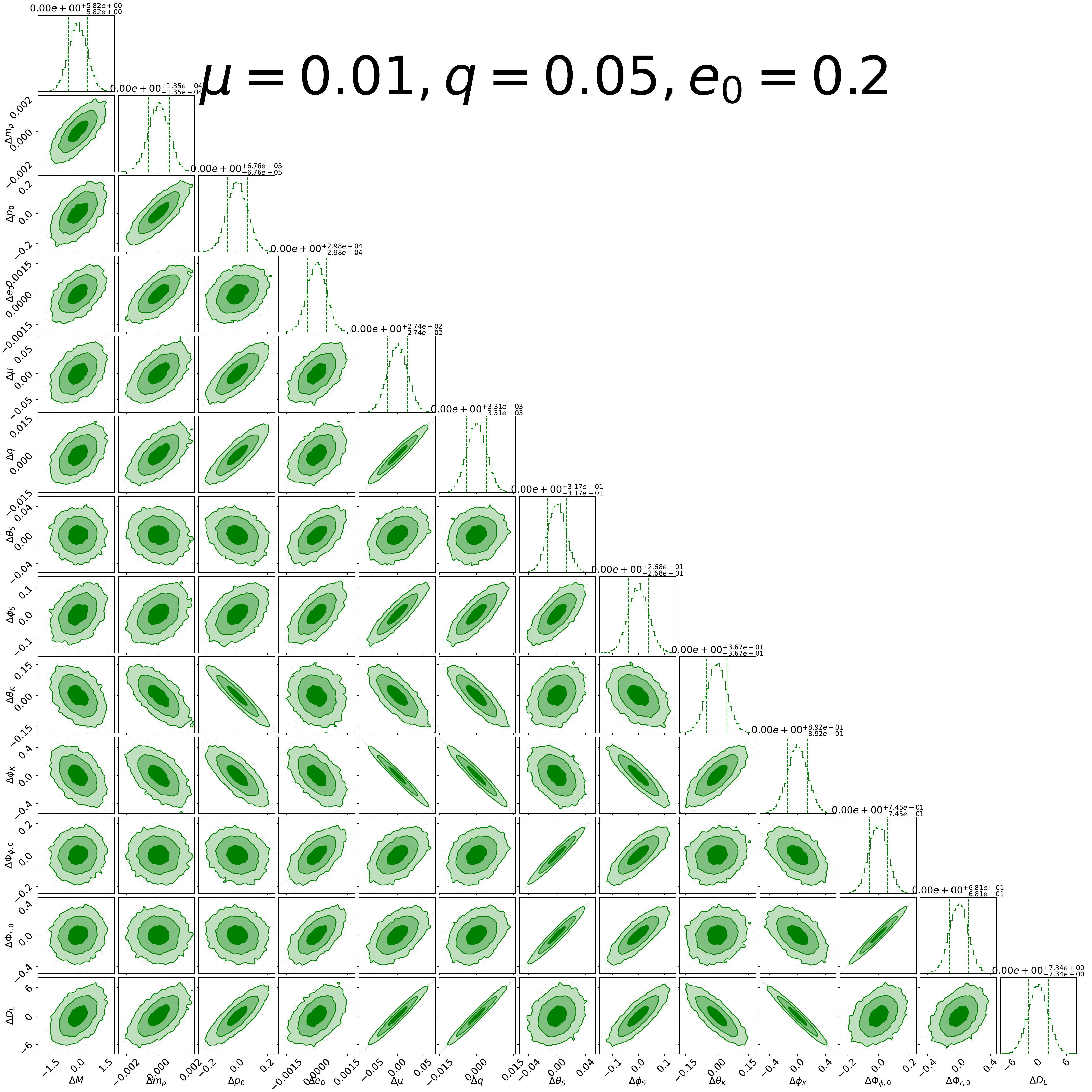}
\caption{Corner plot of the probability distribution for
the masses of EMRIs, initial orbital semi-latus rectum and eccentricity, vector mass, vector charge, directional parameters of source and orbital angular momentum and orbital phases $(M=10^6M_\odot,m_p=10M_\odot,p_0=12.0,e_0=0.2,\mu=0.01, q=0.05, \theta_S=1.0, \phi_S= \theta_K=\phi_K= \Phi_{\phi,0}= \Phi_{r,0}=1.0)$ is showed, which are inferred with two-year observation.  Vertical dashes lines denote to the $1~\sigma$ interval for each source parameter. The contours correspond to 68\%, 95\% and 99\% probability confidence interval.
} \label{fig7:cornerplot:ecc02}
\end{figure*}

\begin{figure*}[htb!]
\centering
\includegraphics[width=0.87\paperwidth]{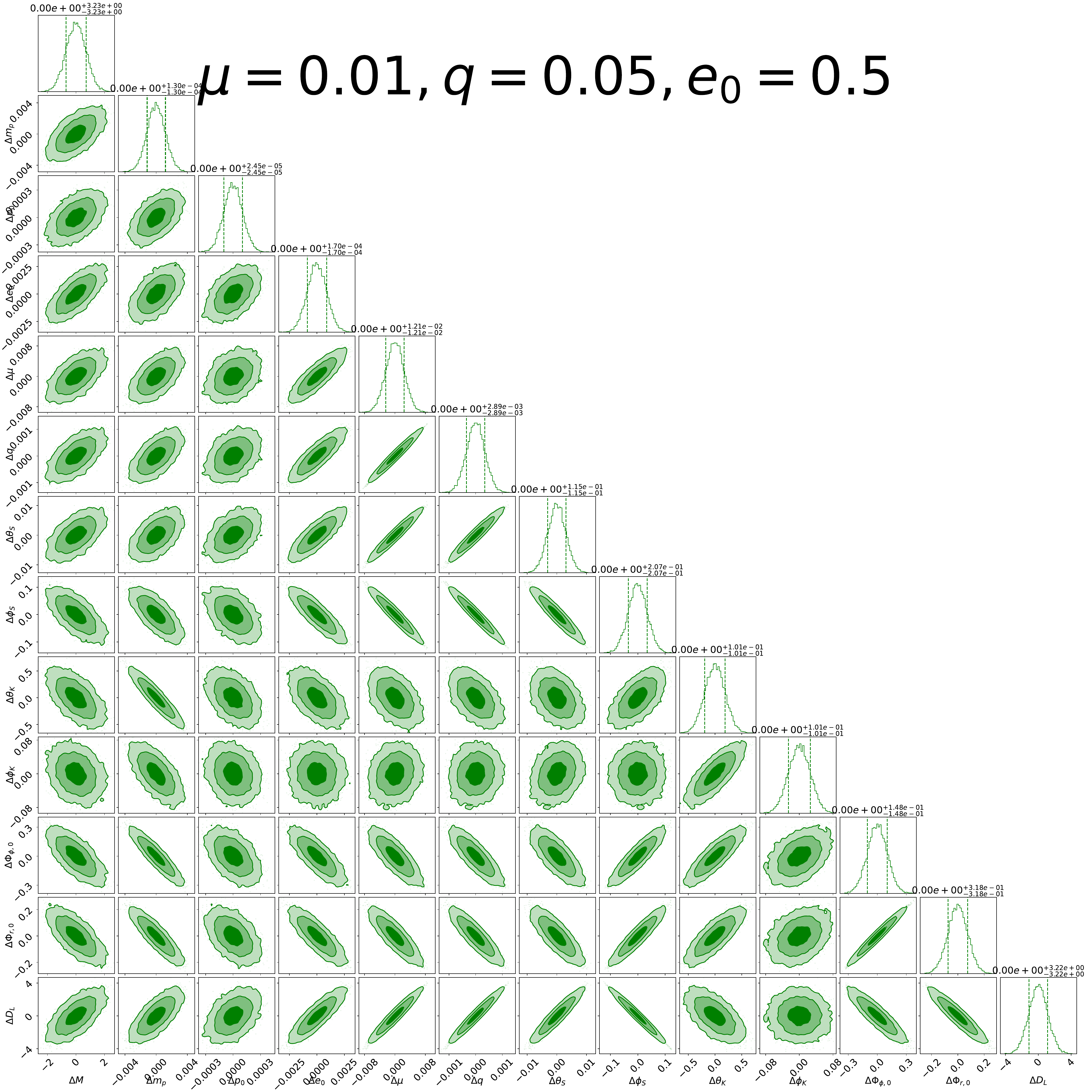}
\caption{Corner plot of the probability distribution for
the masses of EMRIs, initial orbital semi-latus rectum and eccentricity, vector mass, vector charge, directional parameters of source and orbital angular momentum and orbital phases $(M=10^6M_\odot,m_p=10M_\odot,p_0=12.0,e_0=0.5,\mu=0.01, q=0.05, \theta_S=1.0, \phi_S= \theta_K=\phi_K= \Phi_{\phi,0}= \Phi_{r,0}=1.0)$ is showed, which are inferred with two-year observation.  Vertical dashes lines denote to the $1~\sigma$ interval for each source parameter. The contours correspond to 68\%, 95\% and 99\% probability confidence interval.
} \label{fig8:cornerplot:ecc05}
\end{figure*}

We analyze the correlations among the parameters $\boldsymbol{\theta}_i$ using the off-diagonal elements of the covariance matrix in Eq.~\eqref{eq:cov:matrix}, as illustrated in Figs.~\ref{fig7:cornerplot:ecc02} and \ref{fig8:cornerplot:ecc05}. Distinct correlation patterns emerge across the parameter space. Notably, the positive correlation between the primary mass $M$ and the directional angle $\theta_S$ strengthens as the initial eccentricity increases from $e_0=0.2$ to $e_0=0.5$. Similar changing behavior of positive correlations are also observed between $M$ and the parameters $(m_p, p_0, e_0, \mu, q, D_L)$. For more eccentric inspirals, the correlations between $M$ and the phase parameters $(\Phi_{r,0}, \Phi_{\phi,0})$ become strongly positive. Importantly, we identify a robust positive correlation between the Proca mass $\mu$ and several intrinsic parameters $(M, m_p, p_0, e_0, q, D_L)$.
As a result, a better measurement error of intrinsic parameters is useful to constraint on the mass of charge vector field.

In the context of parameter estimation based on the FIM, the measurement uncertainties and correlations among source parameters are obtained from the covariance matrix, defined as the inverse of the FIM. In order to assess the numerical stability of this inversion for a $13\times13$ FIM, we adopt the procedure developed in previous studies~\cite{Speri:2021psr,Maselli:2021men,Zi:2022hcc}. Specifically, we make a perturbation matrix $\mathbf{R}$ of the same dimension as the FIM $\mathbf{\Gamma}$, in which its elements randomly are extracted from a uniform distribution $u \in [-10^{-3},10^{-3}]$. We then compute the inverse of the perturbed matrix $(\mathbf{R}+\mathbf{\Gamma})$ and evaluate the maximum relative deviation between the perturbed and unperturbed covariance matrices, defined as
\begin{equation}
\Delta \mathbf{\Gamma}_d \equiv \max \left( \frac{(\mathbf{R}+\mathbf{\Gamma})^{-1}-\mathbf{\Gamma}^{-1}}{\mathbf{\Gamma}^{-1}} \right).
\end{equation}
Figure~\ref{Fig:CDF} shows the cumulative distribution of $\Delta \mathbf{\Gamma}_d$ for two orbital eccentricities, $e_0 \in \{0.2, 0.5\}$, and for several numerical derivative spacings used in the computation of the FIM for a Proca field with mass $\mu = 0.02$. In both cases, the majority of realizations exhibit maximum relative errors below $0.3\%$, that is over $90\%$ of the population satisfying $\Delta \mathbf{\Gamma}_d \leq 0.3\%$. These results indicate that the covariance matrices obtained from the inversion of the FIM should be numerically stable in our analysis.

\section{Conclusion}\label{conslusion}
In this section, we summarize the main results of the Proca mass correction on the GWs from eccentric EMRIs. As argued before, we begin to compute the Proca and gravitational fluxes using the relativistic method, making generate the adiabatic approximation inspiraling trajectories within the modified \texttt{FEW} model.
During computing the inspiraling orbits in the \texttt{FEW} model, we employ the interpolation method to fast obtain the fluxes according to the relativistic flux data on the rectangle grid. Then we assess the difference of EMRIs signals among the standard GR and the massive or massless vector fluxes by comparing the dephasing and mismatch.
Our study indicates that the inclusion of the orbital eccentricity have indeed the significant influence on the distinguishable of the Proca mass, hence the dephasing and mismatch due to the effect of the vector radiation are also increasing. Considering the initial orbital eccentricity $e_0=0.2$, the imprint of Proca mass on EMRIs quadrupole waveforms can be distinguished within a fractional vector mass
of $\mu\sim10^{-4} ~(\sim1.33\times10^{-20} \rm eV)$ from the standard GR waveforms. When the initial orbital eccentricity is larger up to $e_0=0.4$, the threshold value of the Proca mass discerned by LISA can reach to
$\mu\sim 5\times 10^{-5} ~(\sim6.67\times10^{-21} \rm eV)$.
On the other hand, when comparing the difference of EMRIs waveforms between the Proca field and the massless vector field case, the mismatches would be slightly decrease and the minimum value of the Proca mass with $\mu\sim 5\times 10^{-4} ~(\sim6.67\times10^{-20} \rm eV)$ can be narrowly distinguished. Therefore, the EMRIs fluxes and waveforms imprinted by the lighter Proca field is more sensitive to the orbital eccentricity,
which should be modeled seriously in the relativistic framework.

We finally investigated the capability of LISA to constrain the mass of a massive vector field through parameter estimation of EMRIs. Using the FIM approach, we find that orbital eccentricity plays a crucial role in improving the precision of source parameter estimation. For a Proca field with mass $\mu = 0.01$, the relative uncertainty decreases from $774\%$ to $121\%$ as the orbital eccentricity increases, while for $\mu = 0.02$, LISA can measure the Proca mass with a relative error of $6.54\%$ ($\sim 8.72\times10^{-18}\,\mathrm{eV}$). This improvement is due to that eccentric EMRIs probe deeper into the strong-field regime, encoding richer information about the source parameters. Correlation analysis reveals strong positive dependencies between the Proca mass $\mu$ and other parameters such as $(M, m_p, p_0, e_0, q, D_L)$, suggesting that enhanced accuracy in these parameters directly benefits constraints on $\mu$. Furthermore, numerical stability assessments of the covariance matrix based on perturbations of the FIM following Refs.~\cite{Speri:2021psr,Maselli:2021men,Zi:2022hcc} show that the maximum relative error remains below $0.3\%$ for over $90\%$ of the population, confirming the robustness of our parameter estimation. Overall, our results demonstrate that observations of eccentric EMRIs offer a promising avenue for probing the mass of the Proca field and testing deviations from general relativity in the strong-field regime.

Our analysis is based on the interpolation fluxes in terms of the flux data on the rectangle grid, which essentially exists the interpolation errors. It may be feasible to recede this error using the Chebyshev polynomials as the method developed in Ref.~\cite{Lynch:2023gpu}.
Meanwhile, it would be essential to implement the assessment of measurement error of lighter vector field using the relativistic fluxes and waveforms in the analysis framework of fisher information matrix and Beyesian theorem \cite{Katz:2021yft,Speri:2024qak}.
Realistic EMRIs signal emitted from a rotating MBH, in this scenario
their orbits may be floating due to the mechanism of superradiance as shown in Refs.~\cite{Cardoso:2011xi,Barsanti:2022vvl}, we will work toward the direction in the near future.

\section*{Acknowledgements}
The previous computation on the flux formula derivation and code programming are closely related with the contribution by Chao Zhang in Ningbo University, we thank for his enthusiast help. This work is supported by the National Natural Science Foundation of China with Grants No. 12347140, 12405059 (T.Zi) and  No. 12375049, and Key Program of the Natural Science Foundation of Jiangxi Province under Grant No.20232ACB201008 (F.W.Shu).

%\bibliography{JN1}
%\bibliographystyle{unsrt}
%\bibliographystyle{JHEP}

\providecommand{\href}[2]{#2}\begingroup\raggedright\endgroup

\end{document}